\documentclass[acmsmall,screen]{acmart}
\usepackage{url}
\usepackage{booktabs}
\usepackage{graphicx}
\usepackage{tikz}
\usetikzlibrary{arrows.meta, positioning}
\usepackage{fontawesome5}
\usepackage{todonotes}
\usepackage{stfloats}
\usepackage{placeins}
\usepackage{tcolorbox}
\tcbuselibrary{breakable}
\usepackage{xspace}

\AtBeginDocument{%
  }
\setcopyright{acmlicensed}
\copyrightyear{2026}
\acmYear{2026}
\acmDOI{XXXXXXX.XXXXXXX}
\acmJournal{TOCE}
\acmVolume{0}
\acmNumber{0}
\acmArticle{0}
\acmMonth{0}

\providecommand{\tagpdfsetup}[1]{}
\tagpdfsetup{table/header-rows={1}}

\newcommand{\panelhdr}[1]{%
  \noindent\hspace{-4pt}{\setlength{\fboxsep}{0pt}\colorbox{black}{%
  \parbox{\dimexpr\linewidth+8pt\relax}{\vspace{3pt}\hspace{4pt}%
  \color{white}\scriptsize\bfseries #1\vspace{3pt}}}}}

\newcommand{\tool}{\textsc{Pulla}\xspace}

\makeatletter
\if@ACM@anonymous
  \newcommand{\InstUS}{University~A\xspace}        
  \newcommand{\InstUSname}{University~A\xspace}    
  \newcommand{\InstFI}{University~B\xspace}        

\else
  \newcommand{\InstUS}{the University of Houston\xspace}
  \newcommand{\InstUSname}{University of Houston\xspace}
  \newcommand{\InstFI}{Aalto University\xspace}

\fi
\makeatother

\begin{document}

\title{Trajectories Through Parsons Problems: Making Student Difficulty Patterns Visible to Instructors}

\author{Daniel Prol}
\orcid{0009-0006-6522-4322}
\affiliation{
    \institution{University of Houston}
    \city{Houston}
    \state{Texas}
    \country{USA}
}
\email{dprol@uh.edu}

\author{Juho Leinonen}
\orcid{0000-0001-6829-9449}
\affiliation{
  \institution{Aalto University}
  \city{Espoo}
  \country{Finland}
}
\email{juho.2.leinonen@aalto.fi}

\author{Arto Hellas}
\orcid{0000-0001-6502-209X}
\affiliation{
  \institution{Aalto University}
  \city{Espoo}
  \country{Finland}
}
\email{arto.hellas@aalto.fi}

\author{Saleh Alkhamees}
\orcid{0000-0002-7447-5899}
\affiliation{%
  \institution{University of Houston}
  \city{Houston}
  \state{Texas}
  \country{USA}
  }
\email{saalkham@cougarnet.uh.edu}

\author{Amin Alipour}
\orcid{0009-0008-1935-0605}
\affiliation{
    \institution{University of Houston}
    \city{Houston}
    \state{Texas}
    \country{USA}
}
\email{maalipou@central.uh.edu}

\renewcommand{\shortauthors}{Prol et al.}

\newcommand{\AbstractCategory}[1]{%
  \par\addvspace{.5\baselineskip}%
  \noindent\textbf{#1:}\quad\ignorespaces
}

\begin{abstract}
\AbstractCategory{Background and Context} Existing Parsons problem tools primarily focus on correctness, indicating whether a student solved a problem, but providing limited visibility into the underlying problem-solving process. Without that visibility, instructors cannot see which blocks caused repeated confusion or which distractors were consistently selected.

\AbstractCategory{Objective} We introduce \tool, a Parsons problem tool that instruments programming assignments to capture fine-grained interaction data. We examine what these behavioral traces make visible to instructors, namely how long students take on a problem and how many moves they make, which recurring errors the distractors expose and how the class arrives at them, whether those errors follow a student from one problem to the next, and which students complete the whole assignment without a single error.

\AbstractCategory{Method} \tool records every block arrangement a student submits together with its per-block feedback, and presents per-student attempt histories, activity logs, and AI-assisted difficulty reports to instructors. After a pilot in an upper-division software design course at \InstUS (United States), we deployed it in an online, self-paced introductory Python course at \InstFI (Finland), where 176 students worked on three exception-handling Parsons problems, and analyzed the resulting traces.

\AbstractCategory{Findings} The distractors exposed four recurring errors: misidentifying exception types, returning a value instead of raising an exception, guarding the wrong condition, and placing correct blocks in the wrong order. On the hardest problem 89\% of the students who submitted a wrong block chose the same incorrect guard, and the most common trajectory passed through a failed submission before reaching the correct arrangement. Errors carried over between problems, with a student who made one on the first problem 1.8 to 2.5 times as likely to make it again later, depending on the error. Students who failed before solving a problem spent two to three times longer on it than those who solved it immediately, while those who never solved it spent as little time as the latter. Twenty-seven of the 171 students who submitted to all three problems solved every one at the first attempt, and doing so on the first problem was associated with doing so on the second but not on the third.

\AbstractCategory{Implications} Behavioral traces expose conceptual difficulties that correctness scores hide, including cases in which much of a cohort fails in the same way. We discuss what this means for designing Parsons problems and distractors.
\end{abstract}





\maketitle
\raggedbottom

\section{Introduction}

Parsons problems are exercises in which students reconstruct a program by arranging shuffled code blocks into the correct order~\cite{parsons2006}. They can optionally include distractors that must be excluded~\cite{parsons2006, ericson2022parsons}.
They naturally generate behavioral traces as students solve them, since each block placement, each removal and retry, and each intermediate solution state encodes conceptual decisions that a student makes during problem solving.

Parsons problems are relatively small exercises designed to be completed in short bursts for logical reasoning, structural comprehension, and problem-solving without syntax friction. They can expose common mistakes in students' problem solving and comprehension, and give instructors the evidence to address them.

Parsons problems have been extended in many directions (interaction traces~\cite{helminen2012}, adaptive difficulty~\cite{ericson2018adaptive}, faded blocks~\cite{weinman2021faded}, and mobile accessibility~\cite{kluthe2026}); however, we found no instructor support tool to analyze students' log traces, leaving the problem-solving process invisible to instructors. 
Without visibility into which blocks caused repeated confusion and which distractors were consistently selected, instructors cannot identify which concepts deserve revisiting or which students may benefit from additional support. 

To address this gap, we develop \tool by extending an existing Parsons problem tool with behavioral logging and an instructor-facing dashboard. \tool presents per-student attempt histories and class-level difficulty patterns. The system instruments Parsons assignments to collect fine-grained interaction traces, including the complete sequence of intermediate solution states between the first block placement and final submission. These behavioral traces are designed to enable instructors to identify recurring difficulty patterns and intervene before those patterns compound into deeper misconceptions. We ran a pilot study in an upper-division Java course at \InstUS, and later deployed it in an online, self-paced introductory Python course at \InstFI, where 176 students worked on three exception-handling Parsons problems.
We address four research questions:

\begin{itemize}
    \item[\textbf{RQ1}] How long do students take on a problem and how many moves do they make, and how does this differ between those who solve it at the first attempt, those who solve it after failing, and those who never solve it?
    \item[\textbf{RQ2}] Which distractors carry the failed submissions within each problem, and how does the class arrive at them?
    \item[\textbf{RQ3}] Do difficulty patterns exhibited by a student on one problem reappear in subsequent problems?
    \item[\textbf{RQ4}] How many students complete the whole assignment without a single error, and do they stay on that path from one problem to the next?
\end{itemize}

This paper makes two primary contributions:

\begin{enumerate}
    \item \tool, a Parsons problem tool that turns block-level traces into difficulty reports for instructors.

    \item A trace-based account of how 176 students worked through three problems, covering the time and moves each group spent, the distractors behind the failures, the recurrence of errors across problems, and the students who never fail.
\end{enumerate}

\section{Related Work}
\label{sec:related}

\subsection{Parsons Problems as a Formative Assessment Tool}
Parsons problems have been studied extensively across contexts and skill levels~\cite{du2020review}. Several platforms support their delivery at scale: the \texttt{js-parsons} library\footnote{\url{https://js-parsons.github.io/}} powers web-based courses and was used by \citet{helminen2012} to capture interaction traces in Python problems; Runestone~\cite{ericson2020runestone} embeds them in interactive textbooks with adaptive scaffolding; Epplets~\cite{kumar2018epplets} offers credit based on actions taken to reach a solution; and MobileParsons~\cite{karavirta2012mobileparsons} extends the format to iOS and Android. A recurring design choice is the use of distractors. \citet{harms2016distractors} and \citet{smith2023distractors} found they increase time-on-task by 14--50\% and reduce success rates, so distractor presence shapes the solving process. More recently, CodeTailor~\cite{hou2024codetailor} uses an LLM to generate personalized puzzles from a student's own incorrect submission. The format has also been extended to support adaptive difficulty~\cite{ericson2018adaptive}, faded blocks~\cite{weinman2021faded}, and accessible mobile delivery~\cite{kluthe2026}. Despite these advances, the platforms we reviewed merely report correctness and attempt counts but provide no built-in mechanism for instructors to diagnose how students arrive at their solutions, such as which distractors they select, how often they backtrack, and what difficulties those choices reveal. What these platforms leave unrecorded is the sequence of intermediate arrangements between a student's first block placement and their final submission.

\subsection{Process Data and Instructor-Facing Feedback}
\citet{helminen2012} demonstrated the diagnostic value of interaction traces in Parsons problems, revealing patterns such as backtracking and block cycling that correctness scores cannot expose. \citet{price2017isnap} extended this to block-based programming through iSnap and later proposed ProgSnap2~\cite{price2020progsnap2} as a cross-institutional standard for recording programming events from first interaction to final
submission. \citet{Benedetti2026TheUO} show what is lost when that record is missing. Analyzing written student submissions on algorithmic tasks, they find that when submissions lack explicit solution steps, the reasoning process that produced them remains opaque to analysis.

Process data has also been used to choose what to show to a student. \citet{karavirta2012mobileparsons} offered an optional hint that recalls an earlier visit to the same arrangement and what followed it; \citet{maharjan2020trails} cluster trails of edit distance to the solution and propose feedback derived from them, without implementing or evaluating it; \citet{kumar2023markov} highlights the next incorrect line once the first complete arrangement has been submitted; and \citet{tithi2025guided} derive chunks from traces and pair them with stepwise hints and self-explanation prompts, which improved accuracy on the targeted rules but left the overall posttest score unchanged. Using the history of a session is therefore not new in itself. What varies little across this work is who receives the result, since in these systems it is the student. Among the Parsons studies we reviewed, the only one that names the instructor as a recipient is a two-page abstract in which the tool collects comments written by students rather than a diagnosis derived from their actions~\cite{moloughney2026aat}.

Process data has therefore served as a research instrument and as a source of feedback for students, but not as much as a resource for instructors, and our tool addresses it by exposing each student's complete block-level interaction history within the instructor dashboard.

\subsection{AI-Assisted Analysis of Student Work}
\citet{phung2025} found that expert rubric ratings of AI-generated programming hints tend to overestimate student-perceived usefulness, with most mismatches arising when experts rated a hint as high-quality but students found it unhelpful. A subsequent deployment by the same group~\cite{phung2026} integrated AI-generated hints with an escalation mechanism allowing students to request human instructor feedback when AI support fell short; their results showed that instructor feedback was incorrect or insufficient roughly half the time, even among instructors with domain expertise. \citet{soelch2026icse} integrated LLM-based assessment into a production learning management system, generating rubric-aligned feedback on student UML submissions for instructor review before delivery. Recent work has shown that students rate AI-generated lecture slides comparably to human-authored ones and cannot reliably identify which slides are AI-generated~\cite{leinonen2026aigeneratedslidesgoodstudents}. These approaches take either a student artifact (code, UML) or course content as input and process it without access to the student's problem-solving process. Our approach instead takes a behavioral trace as input and delivers its analysis to the instructor, who remains the decision-maker for any intervention.

\section{System Overview}
\label{sec:system}

\begin{figure*}[t]
  \centering
  \resizebox{\textwidth}{!}{%
  \begin{tikzpicture}[
    arr/.style={-{Stealth[scale=0.9]}, thick, color=black},
    panel/.style={draw=black!60, thin, fill=white, rounded corners=2pt, inner xsep=4pt, inner ysep=0pt},
  ]
\node[font=\normalsize, rotate=90] at (-0.35, 3.5)
  {\faUserGraduate~{\small\bfseries Student}};
\node[font=\normalsize, rotate=90] at (-0.35, -4.0)
  {\faChalkboardTeacher~{\small\bfseries Instructor}};
  \node[panel] (ps1) at (2.3,3.9) {\begin{minipage}[c]{3.8cm}\noindent\hspace{-4pt}{\setlength{\fboxsep}{0pt}\colorbox{black}{\parbox{\dimexpr\linewidth+8pt\relax}{\vspace{3pt}\hspace{4pt}\color{white}\scriptsize\bfseries A -- PROBLEM STATEMENT\vspace{3pt}}}}\\[3pt]\includegraphics[width=3.6cm,keepaspectratio,alt={Screenshot of the problem statement shown to the student}]{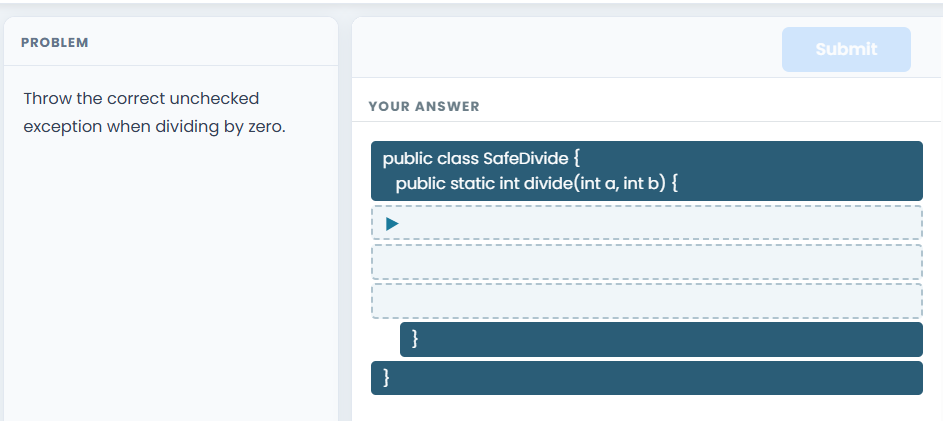}\vspace{3pt}\end{minipage}};%
  \node[panel,right=0.5cm of ps1] (ps2) {\begin{minipage}[c]{3.8cm}\noindent\hspace{-4pt}{\setlength{\fboxsep}{0pt}\colorbox{black}{\parbox{\dimexpr\linewidth+8pt\relax}{\vspace{3pt}\hspace{4pt}\color{white}\scriptsize\bfseries B -- BLOCK BANK\vspace{3pt}}}}\\[3pt]\includegraphics[width=3.6cm,keepaspectratio,alt={Screenshot of the bank of shuffled code blocks, including distractors}]{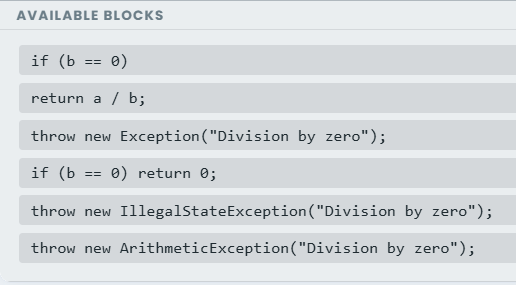}\vspace{3pt}\end{minipage}};%
  \node[panel,right=0.5cm of ps2] (ps3) {\begin{minipage}[c]{3.8cm}\noindent\hspace{-4pt}{\setlength{\fboxsep}{0pt}\colorbox{black}{\parbox{\dimexpr\linewidth+8pt\relax}{\vspace{3pt}\hspace{4pt}\color{white}\scriptsize\bfseries C -- SOLUTION AREA\vspace{3pt}}}}\\[3pt]\includegraphics[width=3.6cm,keepaspectratio,alt={Screenshot of the solution area where the student places blocks}]{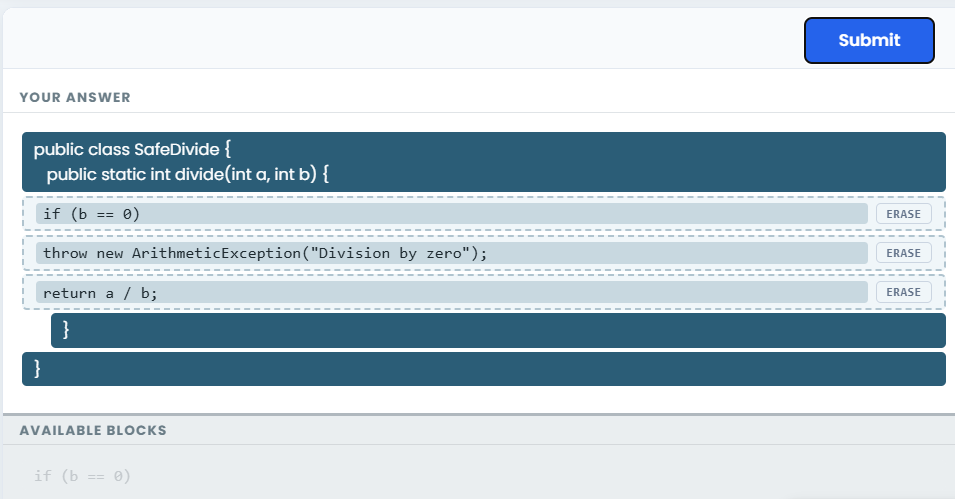}\vspace{3pt}\end{minipage}};%
  \draw[arr] (ps1.east) -- (ps2.west);%
  \draw[arr] (ps2.east) -- (ps3.west);%
  \node[panel] (ps4) at (2.3,0.5) {\begin{minipage}[c]{3.8cm}\noindent\hspace{-4pt}{\setlength{\fboxsep}{0pt}\colorbox{black}{\parbox{\dimexpr\linewidth+8pt\relax}{\vspace{3pt}\hspace{4pt}\color{white}\scriptsize\bfseries D -- FEEDBACK\vspace{3pt}}}}\\[3pt]\includegraphics[width=3.6cm,keepaspectratio,alt={Screenshot of the color-coded per-block feedback after a submission}]{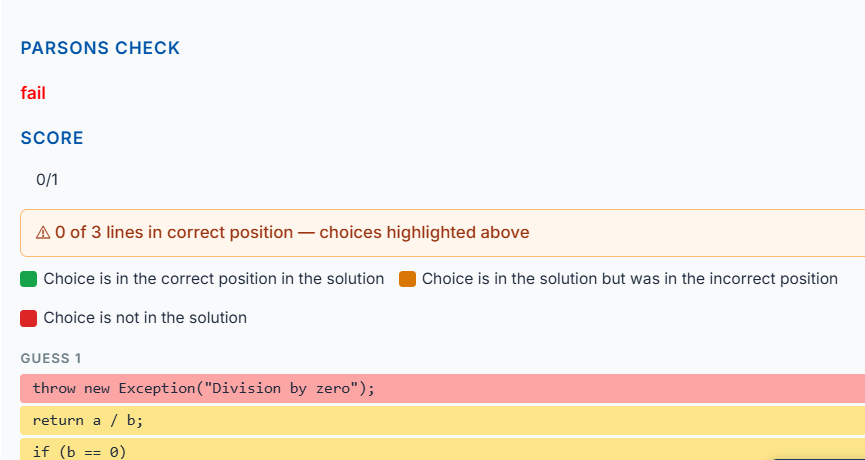}\vspace{3pt}\end{minipage}};%
  \node[panel,right=0.5cm of ps4] (ps5) {\begin{minipage}[c]{3.8cm}\noindent\hspace{-4pt}{\setlength{\fboxsep}{0pt}\colorbox{black}{\parbox{\dimexpr\linewidth+8pt\relax}{\vspace{3pt}\hspace{4pt}\color{white}\scriptsize\bfseries E -- REVISED ATTEMPT\vspace{3pt}}}}\\[3pt]\includegraphics[width=3.6cm,keepaspectratio,alt={Screenshot of the student's revised block arrangement}]{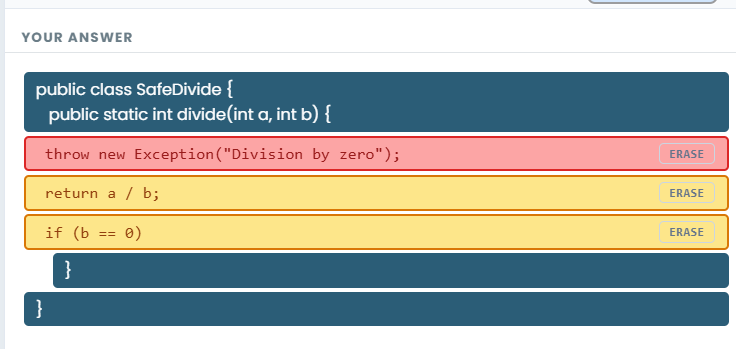}\vspace{3pt}\end{minipage}};%
  \node[panel,right=0.5cm of ps5] (ps6) {\begin{minipage}[c]{3.8cm}\noindent\hspace{-4pt}{\setlength{\fboxsep}{0pt}\colorbox{black}{\parbox{\dimexpr\linewidth+8pt\relax}{\vspace{3pt}\hspace{4pt}\color{white}\scriptsize\bfseries F -- FINAL SUBMISSION\vspace{3pt}}}}\\[3pt]\includegraphics[width=3.6cm,height=2.2cm,keepaspectratio,alt={Screenshot of the final correct submission}]{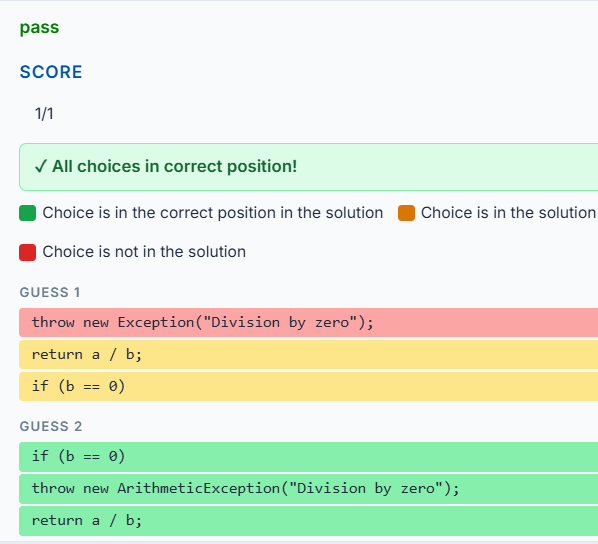}\vspace{3pt}\end{minipage}};%
  \draw[arr] (ps4.east) -- (ps5.west);%
  \draw[arr] (ps5.east) -- (ps6.west);%
  \node[panel] (al) at (1.85,-4.5) {\begin{minipage}[c]{2.9cm}\noindent\hspace{-4pt}{\setlength{\fboxsep}{0pt}\colorbox{black}{\parbox{\dimexpr\linewidth+8pt\relax}{\vspace{3pt}\hspace{4pt}\color{white}\scriptsize\bfseries G -- ACTIVITY LOG\vspace{3pt}}}}\\[3pt]\includegraphics[width=2.7cm,height=2.8cm,keepaspectratio,alt={Screenshot of a student's activity log in the instructor dashboard}]{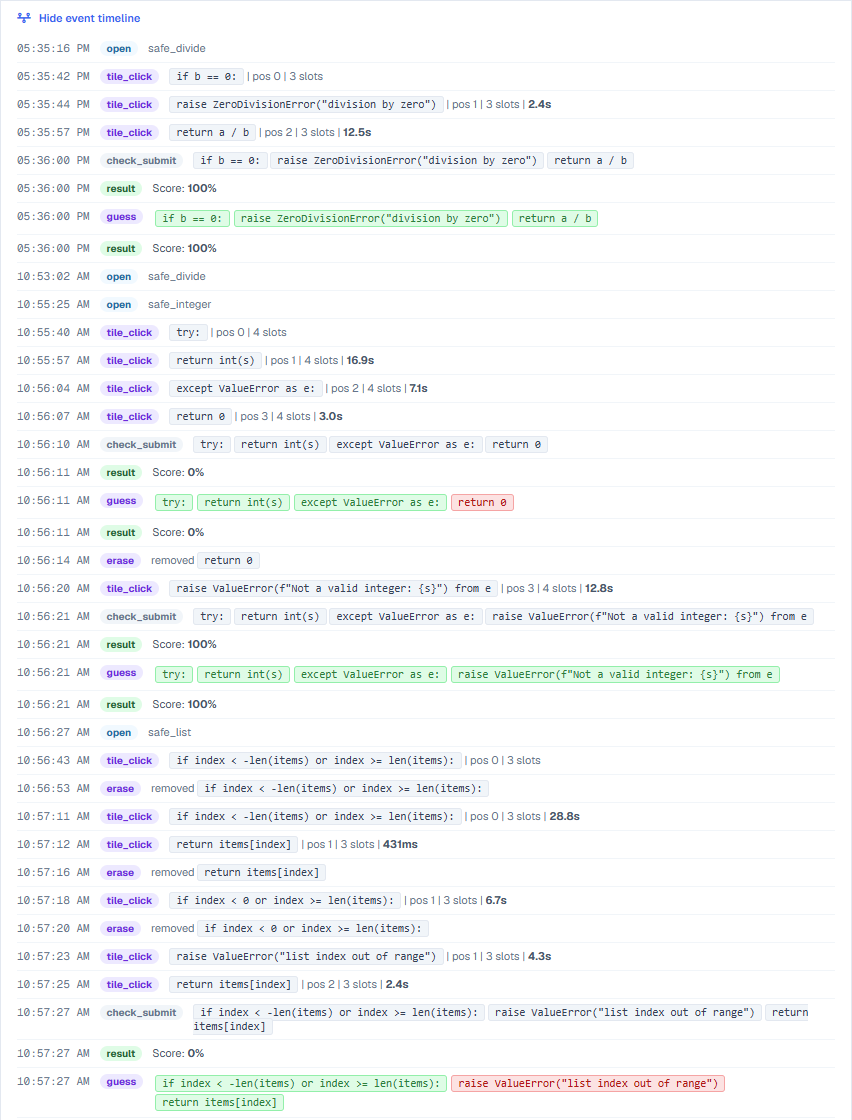}\vspace{3pt}\end{minipage}};%
  \node[panel,right=0.3cm of al] (lk) {\begin{minipage}[c]{2.9cm}\noindent\hspace{-4pt}{\setlength{\fboxsep}{0pt}\colorbox{black}{\parbox{\dimexpr\linewidth+8pt\relax}{\vspace{3pt}\hspace{4pt}\color{white}\scriptsize\bfseries H -- ATTEMPT HISTORY\vspace{3pt}}}}\\[3pt]\includegraphics[width=2.7cm,height=2.8cm,keepaspectratio,alt={Screenshot of a student's attempt history in the instructor dashboard}]{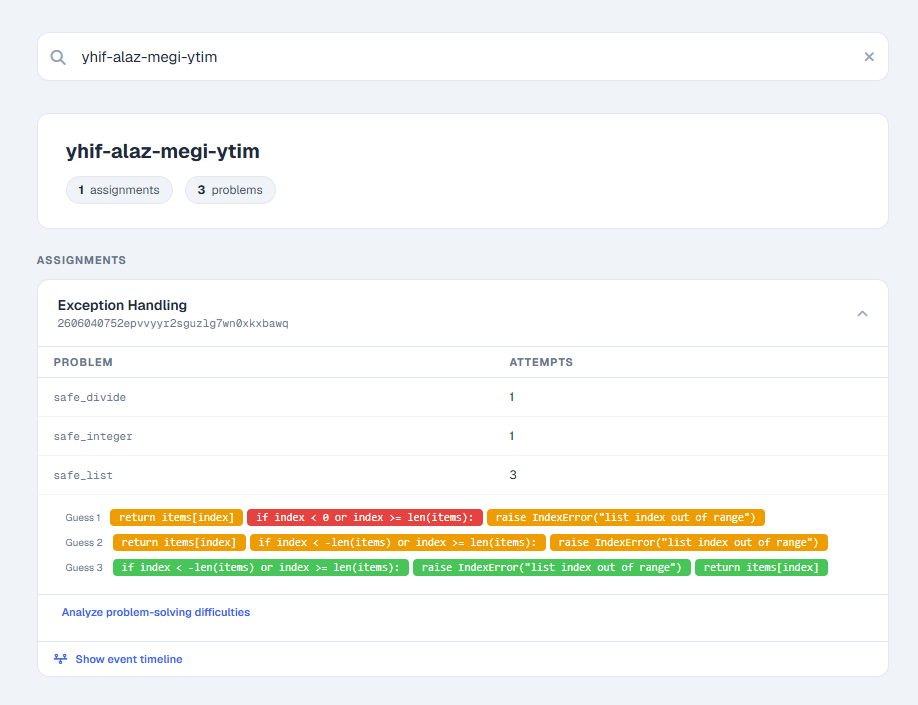}\vspace{3pt}\end{minipage}};%
  \node[panel,right=0.3cm of lk] (mc) {\begin{minipage}[c]{2.9cm}\noindent\hspace{-4pt}{\setlength{\fboxsep}{0pt}\colorbox{black}{\parbox{\dimexpr\linewidth+8pt\relax}{\vspace{3pt}\hspace{4pt}\color{white}\scriptsize\bfseries I -- DIFFICULTIES~{\small\faRobot}\vspace{3pt}}}}\\[3pt]\includegraphics[width=2.7cm,height=2.8cm,keepaspectratio,alt={Generated report naming two of the student's difficulties, with a disclaimer that the analysis may be imperfect}]{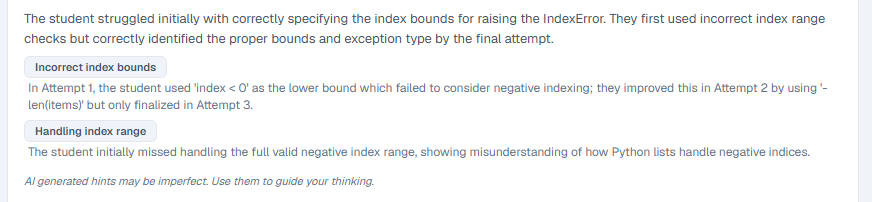}\vspace{3pt}\end{minipage}};%
  \node[panel,right=0.3cm of mc] (sl) {\begin{minipage}[c]{2.9cm}\noindent\hspace{-4pt}{\setlength{\fboxsep}{0pt}\colorbox{black}{\parbox{\dimexpr\linewidth+8pt\relax}{\vspace{3pt}\hspace{4pt}\color{white}\scriptsize\bfseries J -- CLASS SLIDES~{\small\faRobot}\vspace{3pt}}}}\\[2pt]\includegraphics[width=2.7cm,height=1.3cm,keepaspectratio,alt={Generated class slide describing a common difficulty}]{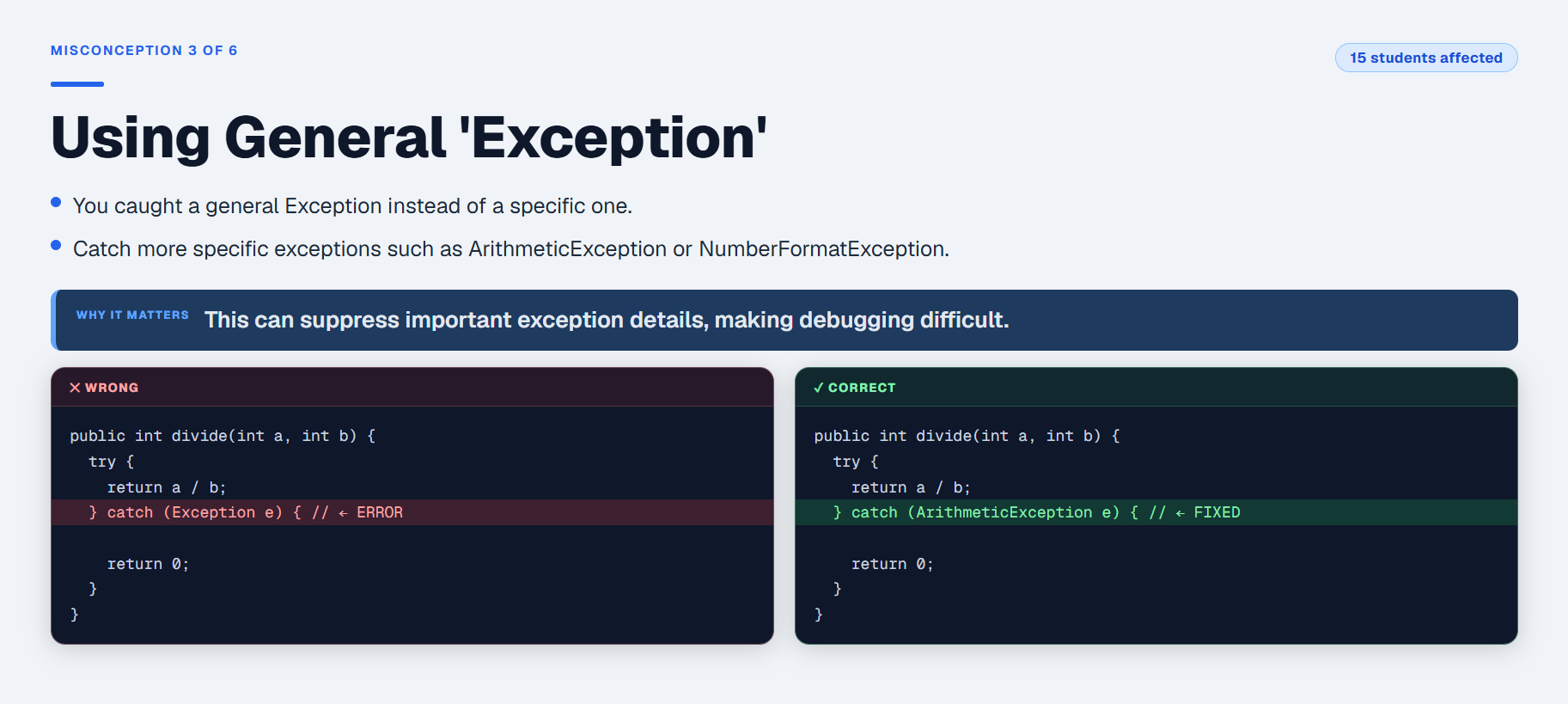}\\[2pt]\includegraphics[width=2.7cm,height=1.3cm,keepaspectratio,alt={Generated class slide summarizing rules to remember}]{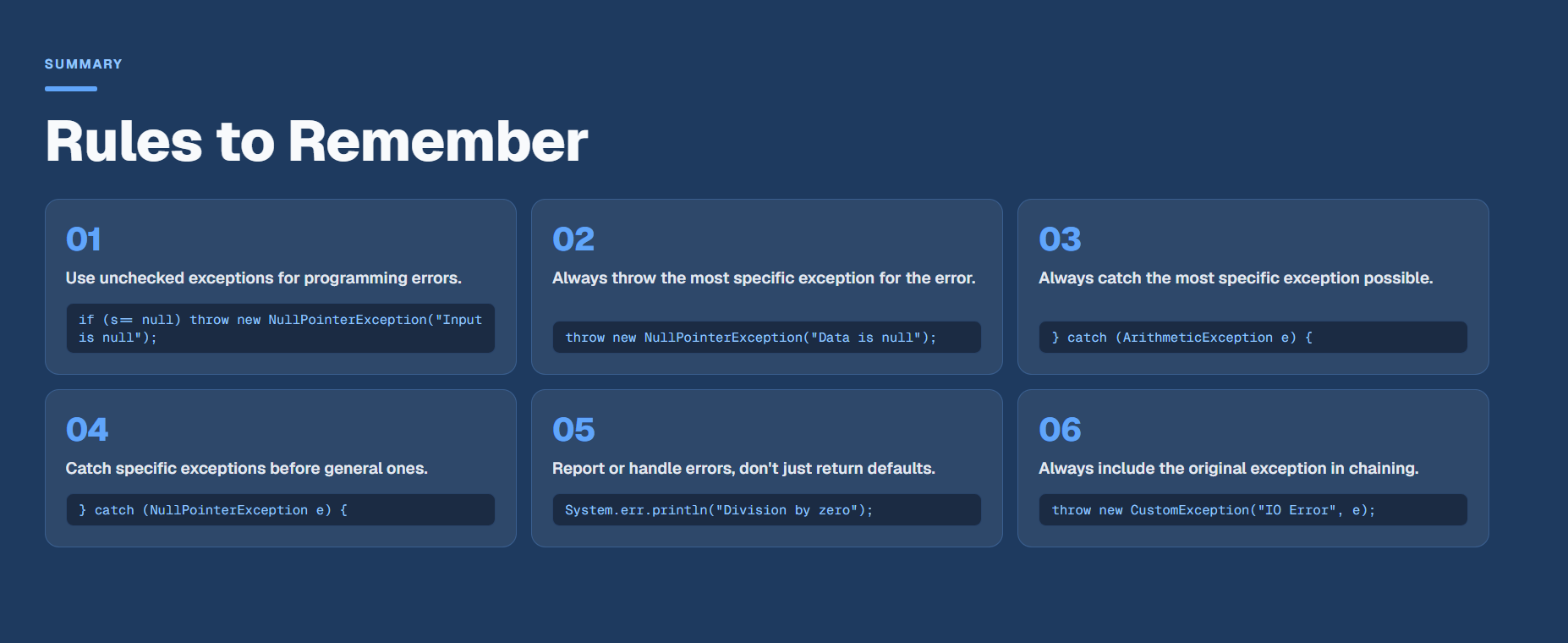}\vspace{3pt}\end{minipage}};%
  \node[draw=black!60, thin, fill=black, rounded corners=2pt, inner sep=5pt, font=\scriptsize\bfseries, text=white] (st) at (4.59,-2.0) {STUDENT PROFILE VIEW};%
  \node[draw=black!60, thin, fill=black, rounded corners=2pt, inner sep=5pt, font=\scriptsize\bfseries, text=white] (gs) at (sl |- st) {SUBMISSIONS};%
  \draw[-,thick,color=black] (ps6.south) -- ++(0,-0.5) coordinate (ps6branch);%
  \draw[arr] (ps6branch) -| (st.north);%
  \draw[arr] (ps6branch) -| (gs.north);%
  \draw[arr] (st.south) -- ++(0,-0.3) -| (al);%
  \draw[arr] (st.south) -- ++(0,-0.3) -| (lk);%
  \draw[arr] (lk.east) -- (mc.west);%
  \draw[arr] (gs.south) -- (sl.north);%
  \end{tikzpicture}}
  \caption{Workflow of \tool. Students solve Parsons problems (A--F) and every arrangement is recorded. Instructors inspect attempt histories~(H) and activity logs~(G), and on demand obtain per-student difficulty patterns~(I) and class-level slides~(J).}
  \Description{System overview with student and instructor panels.}
  \label{fig:overview}
\end{figure*}

\tool  extends CodeCheck~\cite{horstmann_codecheck}, an open-source web-based programming assessment platform, with a behavioral instrumentation layer designed to capture students' problem-solving processes while working on Parsons problems.
Figure~\ref{fig:overview} illustrates the overall workflow of the tool.
Rather than recording only final submission outcomes, the tool captures fine-grained interaction events (Table~\ref{tab:events}) associated with each submission attempt, alongside the per-block feedback returned at each check. These events are persisted for subsequent analysis and together form a behavioral trace, a chronological record of the decisions students make before arriving at a final submission. These intermediate states remain invisible in on-demand feedback systems~\cite{helminen2012}. To support instructional decision-making, the tool provides a web-based instructor dashboard that transforms these traces into actionable insights.
Instructors can inspect per-student attempt histories and activity logs (panels G--H), optionally trigger AI-assisted analysis of problem-solving difficulties for individual students (panel I), and generate a class-level summary of common difficulty patterns as a slide deck that the instructor can present in the browser or download as PowerPoint (panel J).

\begin{table}[!ht]
\caption{Interaction events recorded in the activity log}
\label{tab:events}
\small
\begin{tabular}{@{}l p{9.5cm}@{}}
\toprule
\textbf{Event type} & \textbf{Description} \\
\midrule
problem\_open        & Student navigates to a problem \\
parsons\_tile\_click & Student places a block from the bank into the solution area \\
parsons\_erase       & Student removes a block from the solution area back to the bank \\
check\_submit        & Student submits the current block arrangement for evaluation \\
check\_result        & Score returned for the submitted arrangement \\
parsons\_guess       & Complete block arrangement at submission time with per-block color-coded feedback (green\,/\,yellow\,/\,red); stored as a submission snapshot forming the attempt history \\
\bottomrule
\end{tabular}
\smallskip

{\footnotesize The activity log displays abbreviated labels~(panel~G, Figure~\ref{fig:overview}); the names above reflect the full event types as stored in the database.}
\end{table}

\subsection{Design Rationale and Implementation}

\subsubsection{Client-Side Event Capture}

The tool captures interaction events at the browser level. This design decision is what makes the intermediate states visible, since block placements, removals, and retries are recorded as they happen (Table~\ref{tab:events}), before the student decides to submit. Server-side instrumentation would capture only the final arrangement at each submission, losing the within-attempt behavior that is most diagnostic of conceptual difficulty.

\subsubsection{Human-in-the-Loop Problem-Solving Analysis}

Although the system could deliver AI-generated difficulty reports directly to students, the current deployment positions the instructor as a reviewer first. Figure~\ref{fig:human-in-loop} depicts this workflow. The instructor reviews the student's raw attempt history, optionally invokes the AI-assisted analysis to obtain a labeled difficulty report, and decides how to act on it. The pipeline delivers nothing to students, since its output is written to the instructor dashboard and any use of it, whether a follow-up with the student or an intervention in the next class, is left to the instructor. This follows human-in-the-loop designs for LLM-generated educational feedback~\cite{soelch2026icse}. Three concerns motivate the design. AI-generated feedback can be technically correct yet unhelpful~\cite{phung2025}. AI tutors also tend to shortcut the struggle from which students learn, and embedding pedagogical frameworks in their prompts does not reliably prevent this~\cite{barth2026steering}. Finally, we argue that instructors are better positioned than an automated system to judge whether a detected difficulty pattern is pedagogically meaningful given the course context and students' backgrounds.

\subsubsection{AI Analysis Pipeline}

When an instructor triggers the analysis for a student, the system serializes the student's complete attempt history for the selected assignment, representing each submission as an ordered sequence of block identifiers paired with their per-block color-coded feedback (green, yellow, or red, as described in Section~\ref{sec:parsons-interface}). This structured trace is passed to GPT-4o via the OpenAI API, with a prompt that instructs the model to identify recurring conceptual difficulties evident in the pattern of incorrect placements and distractor selections across attempts. The model returns a short list of labeled difficulty patterns, each with a brief natural-language explanation, displayed to the instructor as the difficulty report (panel~I, Figure~\ref{fig:overview}). For class-level slides~(panel~J), a separate two-stage pipeline first extracts a small set of recurring difficulties from a sample of at most twenty student traces, then classifies every student's attempt history against that shared list, and the six difficulties that most students were mapped to become the deck. Both analyses run on demand. The per-student report is requested from the Student Profile View and the class-level pipeline from the submissions page, and the output of either is stored for the instructor rather than delivered to the student. The same model was used in the pilot and in the main deployment. Section~\ref{sec:perspectives} reports how the instructors experienced these features in practice.

\subsection{Parsons Problem Interface}
\label{sec:parsons-interface}

Students access Parsons problems through a link provided by the course, either in the learning management system (LMS) or as a direct link, which opens a dedicated web-based interface. Upon opening an assignment, students are presented with a problem statement and a bank of shuffled code blocks, including distractors (Figure~\ref{fig:overview}, panels A--B). Students fill the slots by clicking blocks in the bank (panel C). Unlike systems where indentation must be set manually, our tool aligns blocks automatically upon placement, avoiding by design a source of difficulty documented in prior work, since \citet{helminen2012} found that in one assignment three distinct misindentation patterns together accounted for 83\% of solutions. Once all blocks are filled, students submit their arrangement to receive immediate color-coded visual feedback analogous to the Wordle game. Each slot is marked green if the block is in the correct position, yellow if the block belongs somewhere else in the solution, and red if the block is a distractor that does not appear in the correct solution at all (panel~D). Unlike a binary correct/incorrect judgment, this per-block feedback encodes both presence and position simultaneously, allowing students to identify which blocks need repositioning. Crucially, each submission together with its color-coded result is recorded as part of the student's behavioral trace (panel~H), making visible the reasoning path that produced the final submission. Students can then revise their arrangement and submit again (panels E--F).

\subsection{Instructor Interface}

From any assignment, instructors open the Student Profile View (Figure~\ref{fig:overview}, panels G--H) to search for a specific student and inspect their full attempt history and activity log. Instructors can use this view to prepare targeted feedback for students who may need additional support before office hours or tutoring sessions. From the same view, instructors can trigger an automated analysis of problem-solving difficulties that processes the student's attempt history and returns a set of labeled difficulties (panel~I, Figure~\ref{fig:overview}), providing a structured starting point for the conversation. From the general submissions view, instructors can generate a class-level slide deck summarizing common difficulty patterns across the cohort (panel~J, Figure~\ref{fig:overview}). The deck opens in the browser as a presentation and can be downloaded as a PowerPoint file or sent to the instructor by email.

\subsection{Recorded Data}

Both the activity log and attempt history are persisted to a PostgreSQL database. Each row records an event type, a timestamp, and a JSON payload. Interaction events (block placements, removals, submission checks, and problem navigation) form the activity log; Table~\ref{tab:events} lists the six event types captured. Submission snapshots, each recording the complete block arrangement and per-block color-coded feedback, form the attempt history. The tool records identifiers as provided by the course platform. At \InstFI the platform assigns randomly generated pseudonyms natively, and at \InstUS student identifiers were pseudonymized prior to analysis.

\begin{table}[!ht]
\caption{Pilot and main deployment.}
\label{tab:deployments}
\small
\begin{tabular}{@{}lll@{}}
\toprule
                        & \textbf{Pilot (\InstUSname)} & \textbf{Main deployment (\InstFI)} \\
\midrule
Country                 & United States                  & Finland                      \\
Course level            & Upper-division software design & Introductory programming     \\
Student population      & CS majors                      & Non-CS majors                \\
Language                & Java                           & Python                       \\
Format                  & In-person                      & Online, self-paced           \\
Period                  & Spring 2026                    & Summer 2026                  \\
Students                & 68                             & 176                          \\
\bottomrule
\end{tabular}
\end{table}

\begin{figure}[tp]
  \centering
  \begin{tikzpicture}[
    arr/.style={-{Stealth[scale=0.9]}, thick, color=black},
    panel/.style={draw=black!60, thin, fill=white, rounded corners=2pt, inner xsep=4pt, inner ysep=0pt},
  ]
  \node[panel] (pa) {\begin{minipage}[c]{\dimexpr\linewidth-14pt\relax}\panelhdr{A -- ATTEMPT HISTORY: WHAT THE INSTRUCTOR SEES FIRST}\\[3pt]\includegraphics[width=\linewidth,alt={Dashboard table listing the three problems with the number of attempts, the three submitted arrangements for P3 with color-coded blocks, and two links: analyze problem-solving difficulties, and show event timeline}]{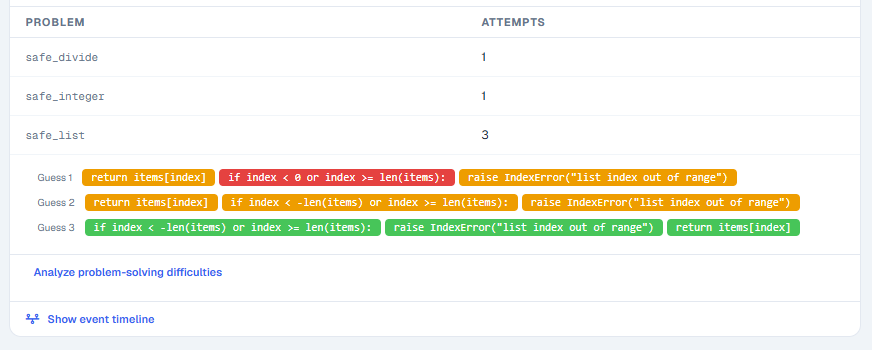}\vspace{3pt}\end{minipage}};%
  \node[panel,below=1.1cm of pa] (pb) {\begin{minipage}[c]{\dimexpr\linewidth-14pt\relax}\panelhdr{B -- EVENT TIMELINE: THE PROCESS BEHIND THE ATTEMPTS}\\[3pt]\includegraphics[width=\linewidth,alt={Timestamped list of events for the same student on the same problem: opening it, placing and removing blocks, and submitting, with the time spent before each placement}]{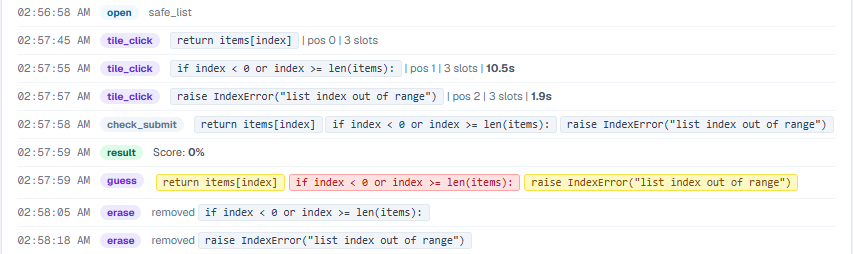}\vspace{3pt}\end{minipage}};%
  \node[panel,below=1.1cm of pb] (pc) {\begin{minipage}[c]{\dimexpr\linewidth-14pt\relax}\panelhdr{C -- DIFFICULTIES, ACCEPTED OR REJECTED BY THE INSTRUCTOR~{\small\faRobot}}\\[3pt]\includegraphics[width=\linewidth,alt={Generated summary of the student's difficulties with two labeled difficulties, incorrect index bounds and handling index range, and a disclaimer that the hints may be imperfect}]{figures/fig2_c_analysis}\vspace{3pt}\end{minipage}};%
  \draw[arr] (pa.south) -- (pb.north);%
  \draw[arr] (pb.south) -- (pc.north);%
  \end{tikzpicture}
\caption{Human-in-the-loop review, for one student on P3. The instructor sees the submitted arrangements (A), may open the event timeline to see the placements and removals between them (B), and may request the AI analysis (C). Each arrow is an optional step, and the analysis is a suggestion the instructor may disregard.}
\Description{Three stacked dashboard screenshots connected by downward arrows. The first is a table of problems and attempts with the submitted arrangements shown as color-coded blocks. The second is a timestamped event timeline of block placements, removals and submissions. The third is a generated summary naming two difficulties, with a disclaimer.}
\label{fig:human-in-loop}
\end{figure}

\section{Method}
\label{sec:method}

\subsection{Study Design}

We evaluated \tool in two stages (Table~\ref{tab:deployments}). A pilot in an upper-division software design course at \InstUS tested the instrumentation, the instructor dashboard, and an initial set of exception-handling problems in classroom use. The main deployment, in an introductory programming course at \InstFI, is the source of all results in Section~\ref{sec:results}. We report the pilot to document how the tool and the problems were developed, not as a second cohort for comparison, since the two courses differ in programming language, course level, student population, and instructional format.

\subsection{Pilot Study}

The pilot ran during Spring 2026 in an upper-division Java course on software design at \InstUS, a large public research university in the United States. Sixty-eight students submitted at least one arrangement, producing 560 submissions. Students solved the problems through the institution's LMS using university credentials, which allowed traces to be linked across assignments; student identifiers were pseudonymized before analysis. The course used Parsons problems on several topics, including design patterns; exception handling was the one topic it shared with the curriculum of the main deployment.

The pilot's exception-handling assignment contained four problems (Table~\ref{tab:pilot}). Because students could submit any subset, the number of students varies by problem, from 64 on \texttt{ExceptionChaining} to 68 on \texttt{SafeDivide}; we report $n=68$ for the assignment as a whole and the per-problem counts in Table~\ref{tab:pilot}.

\begin{table}[t]
\caption{Problems in the pilot, and what became of each in the main deployment.
  Sl./Bl./Dist.\ = slots\,/\,blocks\,/\,distractor blocks;
  \textit{n} = students who submitted at least one arrangement.}
\label{tab:pilot}
\footnotesize
\begin{tabular}{@{}l r r p{5.6cm}@{}}
\toprule
\textbf{Pilot problem (Java)} & \textbf{Sl./Bl./Dist.} & \textit{\textbf{n}} & \textbf{Main deployment} \\
\midrule
\texttt{SafeDivide}          & 3\,/\,6\,/\,3 & 68 & Carried over as P1 \texttt{safe\_divide} \\
\texttt{SafeInteger}         & 4\,/\,7\,/\,3 & 67 & Carried over as P2 \texttt{safe\_integer} \\
\texttt{CheckedVsUnchecked}  & 3\,/\,5\,/\,2 & 66 & Dropped; replaced by P3 \texttt{safe\_list} \\
\texttt{ExceptionChaining}   & 1\,/\,3\,/\,2 & 64 & Dropped \\
\bottomrule
\end{tabular}
\end{table}

Four changes were made between the pilot and the main deployment. First, \texttt{SafeDivide} and \texttt{SafeInteger} were translated to Python as P1 and P2 with the same structure, keeping the same number of slots and blocks and one distractor per error category in the same role (Table~\ref{tab:blocks}). Second, \texttt{CheckedVsUnchecked} was dropped because it turns on Java's distinction between checked and unchecked exceptions, which has no equivalent in Python. Third, \texttt{ExceptionChaining} was dropped because it has a single slot, so a submission is one choice among three blocks rather than an arrangement and there is no solving process to observe. Fourth, the new P3 \texttt{safe\_list} was written to replace them, and it introduces an error category the pilot did not cover, checking the wrong condition (\texttt{if index < 0} instead of \texttt{if index < -len(items)}). The result is a set of three problems of uniform design, each with three distractors, one per error category.

Client-side event logging became available only after the pilot course had ended, so the pilot data consist of submission snapshots only, with no record of the block placements and removals between submissions. The trajectory analyses reported in Section~\ref{sec:results} therefore require the main deployment and cannot be reproduced on the pilot data. What the pilot did establish is that two of the four error categories reported in Section~\ref{sec:results}, returning a value instead of raising an exception and choosing the wrong exception type, were already the dominant failures at submission level, which is why the distractor design was kept.

The pilot instructor described the experience as follows:

\begin{tcolorbox}[
  breakable,
  colback=gray!8, colframe=gray!30,
  arc=2pt, boxrule=0.4pt,
  left=5pt, right=5pt, top=2pt, bottom=2pt
]
\footnotesize
I liked the intuitive interface and the analysis that it provided. Designing new problems was hassle-free. I could use bite-size Parson problems as an in-class exercise that students could do in short amount of time. The summarizing slides were helpful to explain the common mistakes that students make right there in the class. It was also great that they could not copy the solutions from AI, paste the solution in the editor and submit. I felt that I have a quick good estimate of their skill levels.
\end{tcolorbox}

\subsection{Main Deployment}

The main deployment took place in an introductory Python course for non-CS majors at \InstFI in Finland, delivered online and self-paced. Students opened the problems through a direct link without institutional authentication, and the platform assigned each of them a randomly generated pseudonym, so no personal identifiers were recorded. The pseudonym is stable across the three problems, so a student's work can be followed from one to the next without being identifiable, and 179 of the 182 students who placed a block did so on all three problems. This is what makes the comparisons across problems in Sections~\ref{sec:rq3} and~\ref{sec:rq4} possible. We analyzed all activity on the exception-handling assignment during Summer 2026, in which 194 students opened the assignment, 182 placed at least one block, and 176 submitted at least one arrangement. Together they produced 819 submitted arrangements and 3{,}675 block placements and removals, a mean of 20 block moves per student across the three problems.

\subsection{Materials}

The assignment comprised three Parsons problems on exception handling (Table~\ref{tab:problems}). In each, students arranged blocks into a Python function that signals an error condition correctly. Every distractor corresponds to one of three errors: returning a value instead of raising an exception, raising an imprecise or wrong exception type, or checking the wrong condition (Table~\ref{tab:blocks}). Problems were presented in order (P1--P3), but students could move on without solving the current one.

\begin{table}[t]
\caption{Parsons problems in the main deployment.}
\label{tab:problems}
\begin{tabular}{@{}l l p{7.5cm}@{}}
\toprule
\textbf{ID} & \textbf{Sl./Bl./Dist.} & \textbf{Description} \\
\midrule
P1 Safe Divide      & 3\,/\,6\,/\,3 & Raise the correct exception for division by zero \\
P2 Safe Integer     & 4\,/\,7\,/\,3 & Raise the correct exception for invalid string input \\
P3 Safe List Access & 3\,/\,6\,/\,3 & Raise the correct exception for an out-of-range index \\
\bottomrule
\end{tabular}
\smallskip

{\footnotesize Sl./Bl./Dist.\ = slots\,/\,blocks\,/\,distractor blocks, as in Table~\ref{tab:pilot}.}
\end{table}

\begin{table}[t]
\caption{Block sets. Distractors are labeled with the error they embody.}
\label{tab:blocks}
\footnotesize
\begin{tabular}{@{}l l p{3.4cm}@{}}
\toprule
& \textbf{Block} & \textbf{Role} \\
\midrule
\multicolumn{3}{@{}l}{\textit{P1} \quad \texttt{def divide(a, b):}} \\
B1 & \texttt{if b == 0:} & correct \\
B2 & \texttt{if b == 0: return 0} & return instead of raise \\
B3 & \texttt{raise ZeroDivisionError("division by zero")} & correct \\
B4 & \texttt{raise RuntimeError("division by zero")} & exception type \\
B5 & \texttt{raise Exception("division by zero")} & exception type \\
B6 & \texttt{return a / b} & correct \\
\midrule
\multicolumn{3}{@{}l}{\textit{P2} \quad \texttt{def parse(s):}} \\
B1 & \texttt{try:} & correct \\
B2 & \texttt{return int(s)} & correct \\
B3 & \texttt{return 0} & return instead of raise \\
B4 & \texttt{except ValueError as e:} & correct \\
B5 & \texttt{except Exception as e:} & exception type \\
B6 & \texttt{raise ValueError(f"Not a valid integer: \{s\}") from e} & correct \\
B7 & \texttt{raise RuntimeError("Not a valid integer")} & exception type \\
\midrule
\multicolumn{3}{@{}l}{\textit{P3} \quad \texttt{def get\_item(items, index):}} \\
B1 & \texttt{if index < -len(items) or index >= len(items):} & correct \\
B2 & \texttt{if index < -len(items) or index >= len(items): return None} & return instead of raise \\
B3 & \texttt{if index < 0 or index >= len(items):} & wrong guard condition \\
B4 & \texttt{raise IndexError("list index out of range")} & correct \\
B5 & \texttt{raise ValueError("list index out of range")} & exception type \\
B6 & \texttt{return items[index]} & correct \\
\bottomrule
\end{tabular}
\end{table}

\subsection{Measures and Analysis}
\label{sec:measures}

We reconstructed each student's solving session from the event logs (Table~\ref{tab:events}) and the submission snapshots. For each problem, we report the number of students who submitted at least one arrangement ($n$), the mean number of submitted arrangements, the percentage of students whose first submission was correct, and the percentage of submitted arrangements containing at least one distractor. Error categories are computed over the students with at least one such arrangement (Err.~$n$), where a student counts toward a category if any of their submissions contained a distractor of that category, so categories can overlap. Exception type confusion covers an imprecise or wrong exception type; return instead of raise covers distractors that return a value; wrong condition covers a guard that does not delimit the error case correctly; and wrong order covers correct blocks submitted in the wrong position. We report the wrong condition separately rather than folding it into either neighboring category. The block raises the same exception as the correct one, so it is not a confusion about \emph{which} exception to raise, and it is a distractor rather than a correct block out of place, so it is not an ordering error. Only P3 includes such a distractor. A revisit is a submission identical to an earlier submission by the same student on the same problem.

To summarize trajectories at the cohort level, we built one first-order Markov model per problem from all students who placed at least one block (179 in P3), following \citet{kumar2023markov}, who analyzes Parsons puzzle solving with a first-order Markov transition matrix over the lines a student acts on, bounded by dummy start and end states. Our states describe the whole solution area rather than a single line. A state is the arrangement in the solution area after one event, or a failed or passed submission, and the model is bounded by a non-emitting start state and an absorbing end state. 
Writing $n(s \rightarrow s')$ for the number of observed transitions from state $s$ to state $s'$, we take the maximum likelihood estimate $\hat{P}(s' \mid s) = n(s \rightarrow s') / \sum_{u} n(s \rightarrow u)$, with no smoothing. The map annotates each state $s$ with the Shannon entropy of that distribution, $H(s) = -\sum_{s'} \hat{P}(s' \mid s) \log_2 \hat{P}(s' \mid s)$, in bits, where the sum runs over the successors actually observed from $s$. 
Low entropy suggests predictability of students' next actions given the current state, while higher entropy suggests higher level of stochasticity, hence less predictability. 
For instance, in P3,  the state holding the incorrect guard (B3) and \texttt{IndexError} (B4) with the third slot still empty was entered 88 times, and the next event completed it with the return statement (B6) in 84 of those, the remaining 4 splitting over two other blocks, which gives $H = 0.31$ bits against the $\log_2 3 = 1.58$ that three equally likely successors would give. The start state, entered once by each of the 179 students and followed by six different first blocks, gives 1.73 bits. Both values are printed on the corresponding nodes of Figure~\ref{fig:cohort-map}. Because it is computed from estimated rather than known probabilities, $H(s)$ is biased downward at states with few visits, and the map therefore reports the visit count $n$ next to $H$ for every state. Entropy over transition probabilities has been used in the same way to characterize programming behavior from log data~\cite{mai2023entropy}. The map draws a transition as a solid arrow when its probability is at least 15\%, and as a dashed arrow when it is below 15\% but occurred at least 10 times. A state is drawn if it accounts for at least 3\% of all transitions, was visited at least 10 times, or is the target of an arrow that meets either criterion. The predictable path consists of the states reachable from the start through solid arrows only; all other states, and the arrows that touch them, are drawn in grey. A step is \emph{predictable} when at least $\tau=15\%$ of the students in that state took it.

Timing comes from the same event log. Time on a problem is the sum of the intervals between consecutive events in that student's session, discarding any interval longer than ten minutes as time spent away from the tool; the time between moves is the median of those intervals. Because sessions are short and a few are very long, we report medians and quartiles alongside means and sample standard deviations. On each problem we group students into those whose first submission was correct, those who failed at least once and then solved the problem, and those who never solved it.

Comparisons between two groups of students are reported as a risk ratio, the percentage in one group divided by the percentage in the other, with Fisher's exact test rather than a $\chi^2$ test because some expected cell counts fall below five. Where a family of related tests is run, one per error category, we report both the raw $p$ and the value after Holm's correction across that family. Confidence intervals on risk ratios use the log method. Where we compare one such comparison against another, we use a permutation test with 20{,}000 relabelings~\cite{ernst2004permutation}, which keeps each student's two outcomes together.

\subsection{Ethics}

The data collection followed the ethical guidelines of the university where the data was collected, which do not require a separate ethics review when the data are anonymous. In the main deployment students were identified only by the pseudonym the platform generates, so no personal identifiers were recorded; in the pilot, identifiers were pseudonymized before analysis.

\section{Results}
\label{sec:results}

\subsection{RQ1: How Students Work Through a Problem}
\label{sec:rq1}

The event log records when every block was placed and removed. Every placement and removal carries a timestamp (panel~B, Figure~\ref{fig:human-in-loop}), so the time a student spent between one move and the next can be measured. On each problem, we divided the students into three groups. The first group submitted a correct arrangement on the first attempt, the second failed at least once before solving the problem, and the third never solved it. Table~\ref{tab:time} reports how long each group spent, how much time passed between their moves, and how many blocks they placed and removed.

Students who failed before solving the problem spent the longest on it, between two and three times the median of those who solved it immediately. Students who never solved it, however, spent almost as little time as those who solved it on the first attempt, with medians of 18 seconds against 15 on P3, 41 against 44 on P2, and 48 against 29 on P1. Students who never solved the problem acted fastest, every 2.3 to 2.8 seconds, and removed almost nothing, a median of one removal on P1 and none on P2 and P3. Those who solved the problem at the first attempt were the slowest between moves, at 3.8 to 6.7 seconds, with a median of zero removals.

The long sessions belong to the students who failed before solving the problem, who placed five to seven blocks where the solution needs three or four, and removed two or three. They therefore look like productive struggle, in the sense \citet{warshauer2015} gives the term, although the traces cannot show whether the struggle led to learning. Students who solved at the first attempt placed the minimum number of blocks the solution requires and removed none, but took the longest over each move, between 3.8 and 6.7 seconds, which is consistent with thinking before acting. Students who never solved the problem built sessions of a similar size, three or four blocks with almost no removals, but moved through them fastest of all. Their sessions are short because they stop early, not because they go straight to the answer.

\begin{table}[!ht]
\caption{Time and activity per problem, by what happened to the student on it. Seconds on the problem are the sum of the gaps between consecutive events, excluding gaps over ten minutes. Md = median, Q1 and Q3 the first and third quartiles; Moves is the interval between consecutive events, not the total time divided by the number of moves.}
\label{tab:time}
\footnotesize
\setlength{\tabcolsep}{4.5pt}
\begin{tabular}{@{}l l r rrrrr r rr@{}}
\toprule
& & & \multicolumn{5}{c}{\textit{Seconds on the problem}} & \textit{Moves (s)}
& \textit{Blocks} & \textit{Removals} \\
\cmidrule(lr){4-8}
& \textbf{Result} & \textit{n} & Mean & SD & Q1 & Md & Q3 & Md & Md & Md \\
\midrule
P1 & Correct first attempt & 133 &  49 &  81 & 16 & 29 &  46 & 4.8 & 3 & 0 \\
   & Failed, then solved   &  35 & 101 &  85 & 53 & 83 & 105 & 3.6 & 6 & 3 \\
   & Never solved          &   7 &  68 &  72 & 26 & 48 &  71 & 2.6 & 4 & 1 \\
\midrule
P2 & Correct first attempt &  98 &  76 &  97 & 27 & 44 &  97 & 6.7 & 4 & 0 \\
   & Failed, then solved   &  62 & 132 & 101 & 59 & 91 & 188 & 4.0 & 7 & 3 \\
   & Never solved          &  13 &  70 &  85 & 23 & 41 &  93 & 2.8 & 4 & 0 \\
\midrule
P3 & Correct first attempt &  49 &  36 &  68 & 11 & 15 &  41 & 3.8 & 3 & 0 \\
   & Failed, then solved   & 108 &  72 &  62 & 31 & 52 &  87 & 3.6 & 5 & 2 \\
   & Never solved          &  16 &  30 &  37 &  8 & 18 &  30 & 2.3 & 3 & 0 \\
\bottomrule
\end{tabular}
\end{table}

\subsubsection{Individual Trajectories}
\label{sec:loops}

\begin{figure}[tp]
  \centering
  \includegraphics[width=0.78\linewidth,alt={State-transition map of one student's session on problem P1, from an empty solution area through a failed submission to a correct one}]{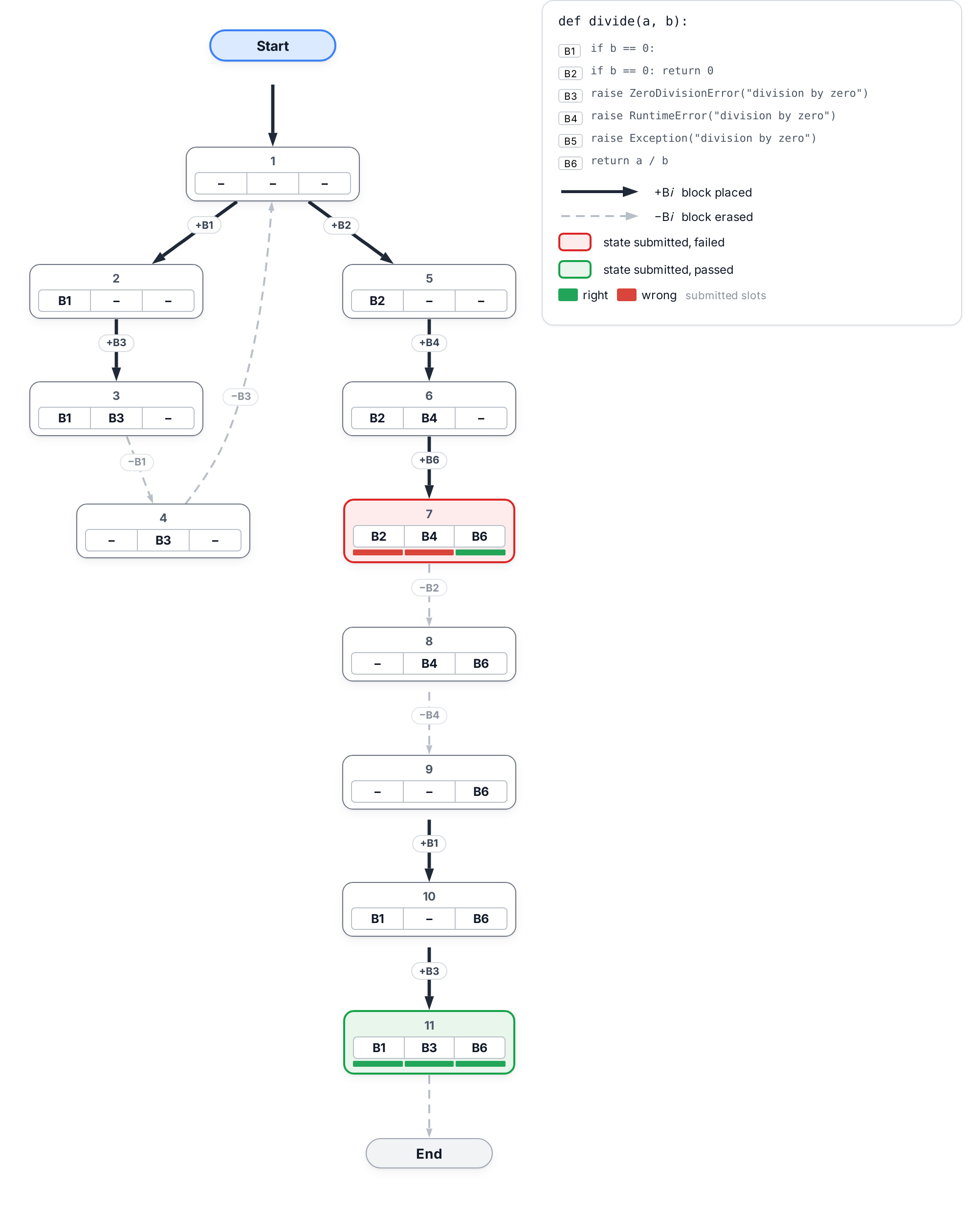}
  \caption{One student's session on P1. Nodes are arrangement states, numbered in order of first appearance, so a state the student returns to keeps its number. Black arrows place a block ($+$B$i$), dashed grey arrows remove one ($-$B$i$), and a red or green node is a state submitted and failed or passed, with its per-block feedback below the slots.}
  \Description{A vertical graph of eleven numbered arrangement states. From the empty solution area (1), the student places B1 (2) and B3 (3), removes B1 (4) and B3, leaving the solution area empty. The student then places B2 (5), B4 (6), and B6 (7); the submission at state 7 fails, with B2 and B4 marked red. The student removes B2 (8) and B4 (9), places B1 (10) and B3 (11), and the submission at state 11 passes with all three blocks green. A block bank for B1 to B6 and a legend appear on the right.}
  \label{fig:state-map}
\end{figure}

Table~\ref{tab:time} says how long a session lasted, not what happened inside it. Figure~\ref{fig:state-map} traces one student's session on P1 as a state-transition map, in which each node is the arrangement after an event, numbered in order of first appearance, and each arrow is a block placement ($+$) or removal ($-$). The student began with the correct guard and exception (states 2--3), then removed both blocks and cleared the solution area (state 1). The second attempt combined two distractors, the guard that returns 0 and \texttt{RuntimeError} (B2 and B4), and the submission failed (state 7). The per-block feedback marked those two blocks red; the student removed exactly those two, placed the correct blocks in their place, and the next submission passed (state 11). The session thus contains a full reset and an abandoned, correct partial arrangement, neither of which is visible in the student's two submissions or final score.

Between 1.7\% and 2.9\% of sessions per problem contained a revisit, well below the 21\% to 33\% that \citet{helminen2012} reported for comparable Parsons tasks.

\subsection{RQ2: Distractors Behind Failed Submissions}
\label{sec:rq2}

Table~\ref{tab:combined} reports how often each of the four error categories appeared. It counts students, over those who submitted at least one arrangement containing a distractor, and groups the individual blocks into categories. The categories are not mutually exclusive, since one submission can raise the wrong exception and also place the blocks in the wrong order, so a student may count toward several of them.

\begin{table*}[t]
\centering
\caption{Problem difficulty and error categories in the main deployment.
  \textit{n} = students who submitted;
  Avg.\ Att.\ = mean submitted arrangements per student;
  1st~Att.\ = correct on first submission;
  Incorr.\ = \% of submitted arrangements containing at least one distractor.
  Error-category percentages are computed over the students with at least one such arrangement
  (Err.\ \textit{n}) and can overlap.}
\label{tab:combined}
\footnotesize
\setlength{\tabcolsep}{4.5pt}
\begin{tabular}{@{}l r rrrr rrrr@{}}
\toprule
& & \multicolumn{4}{c}{\textit{Submissions}}
  & \multicolumn{4}{c}{\textit{Error categories}} \\
\cmidrule(lr){3-6}\cmidrule(l){7-10}
\textbf{Problem} & \textit{n}
  & \textbf{Avg.} & \textbf{1st} & \textbf{Err.} & \textbf{Incorr.}
  & \textbf{Exc.} & \textbf{Ret.\ not} & \textbf{Wrong} & \textbf{Wrong} \\
  & & \textbf{Att.} & \textbf{Att.} & \textit{\textbf{n}} &
  & \textbf{type} & \textbf{raise} & \textbf{cond.} & \textbf{order} \\
\midrule
P1                            & 175 & 1.27 & 76.0\% &  38 & 21\% & 76\% & 42\% & --- & 24\% \\
P2                            & 173 & 1.56 & 56.6\% &  65 & 33\% & 75\% & 43\% & --- & 58\% \\
P3$^\dagger$                  & 173 & 1.88 & 28.3\% & 123 & 49\% & 27\% & 21\% & 89\% & 20\% \\
\bottomrule
\end{tabular}
\smallskip\\
\raggedright\footnotesize
$^\dagger$Hardest problem.
Exc.\ type = exception type confusion;
Ret.\ not raise = \texttt{return} value instead of raising an exception;
Wrong cond.\ = a guard that does not delimit the error case correctly, a distractor only P3 includes (---);
Wrong order = correct blocks submitted in the wrong position.
\end{table*}

Table~\ref{tab:failed} breaks the same failures down by submission and by individual block. A submission can contain more than one distractor, so the percentages within a problem sum to more than 100\%.

One distractor dominates in P3 and none does in P1 or P2. In P3, \texttt{if index < 0} appears in 70\% of the 168 failed submissions, more than three times the next block. In P1 the three distractors take 45\%, 30\%, and 26\% of 53 failed submissions, and in P2 37\%, 30\%, and 24\% of 109.

\begin{table}[!ht]
\caption{Distractors in failed submissions. For each problem, the number and percentage of failed submissions containing each distractor block. A submission can contain several distractors, so the percentages do not sum to 100\%.}
\label{tab:failed}
\footnotesize
\setlength{\tabcolsep}{5pt}
\begin{tabular}{@{}l l r r@{}}
\toprule
\textbf{Problem} & \textbf{Block in the failed submission} & \textit{n} & \textbf{\%} \\
\midrule
\multicolumn{4}{@{}l}{\textit{P1} \quad 53 failed submissions} \\
& B5 \quad \texttt{raise Exception}                & 24 & 45\% \\
& B2 \quad \texttt{if b == 0: return 0}            & 16 & 30\% \\
& B4 \quad \texttt{raise RuntimeError}             & 14 & 26\% \\
\midrule
\multicolumn{4}{@{}l}{\textit{P2} \quad 109 failed submissions} \\
& B5 \quad \texttt{except Exception as e:}         & 40 & 37\% \\
& B3 \quad \texttt{return 0}                       & 33 & 30\% \\
& B7 \quad \texttt{raise RuntimeError}             & 26 & 24\% \\
\midrule
\multicolumn{4}{@{}l}{\textit{P3} \quad 168 failed submissions} \\
& B3 \quad \texttt{if index < 0 or index >= len(items):} & 118 & 70\% \\
& B5 \quad \texttt{raise ValueError}                & 37 & 22\% \\
& B2 \quad guard returning \texttt{None}            & 33 & 20\% \\
\bottomrule
\end{tabular}
\end{table}

\begin{figure}[tp]
  \centering
  \includegraphics[width=\linewidth,alt={Cohort predictability map for problem P3, aggregating 179 student trajectories}]{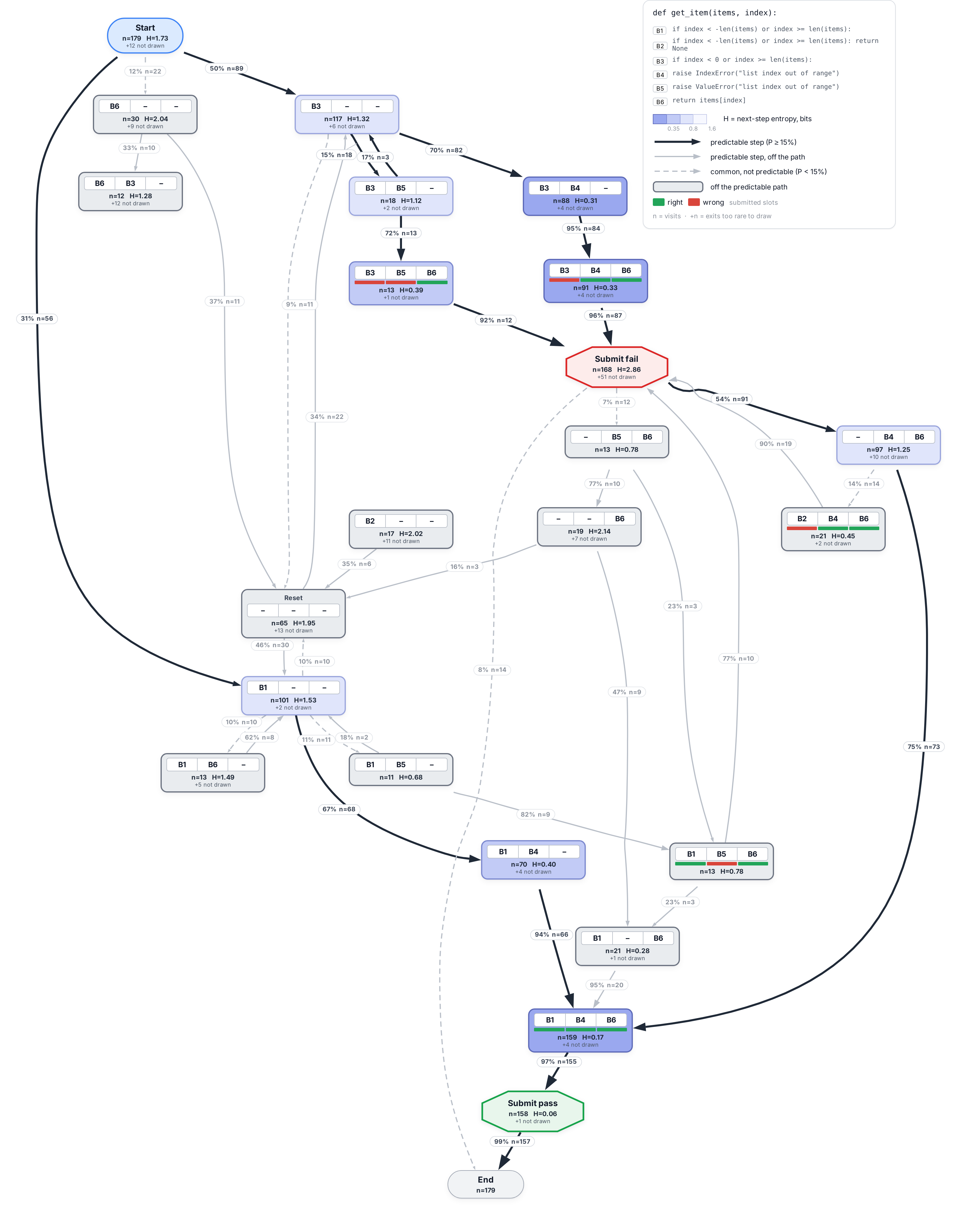}
  \caption{Cohort predictability map for P3, over the trajectories of 179 students. Nodes are arrangement states (Table~\ref{tab:blocks}), with their visits $n$ and next-step entropy $H$; their shading encodes $H$ on the predictable path, from dark for one dominant continuation to light for a decision point, and states off that path are grey. Arrow width grows with the probability of the step. Thresholds are those of Section~\ref{sec:measures}.}
  \Description{A directed graph of about two dozen arrangement states between a start node at the top and an end node at the bottom. Thick black arrows form two main routes: one from the start through B3, B3 B4, and B3 B4 B6 to a failed submission, then through the state with only B4 and B6 to B1 B4 B6 and a passed submission; the other from the start through B1 and B1 B4 to B1 B4 B6. A smaller black branch passes through B3 B5 to a failed submission. Grey nodes, including the state in which the solution area is empty again, sit off these routes and connect through grey and dashed arrows. A legend with the block bank and the visual encodings appears at the top right.}
  \label{fig:cohort-map}
\end{figure}

The trajectories say how the class arrives at that distractor and what it does afterwards, which is what an instructor would act on. Figure~\ref{fig:cohort-map} aggregates the whole cohort on P3 as a first-order Markov model over arrangement states, as described in Section~\ref{sec:measures}.

Two routes carry most of the 179 trajectories. Half of the students opened with the incorrect guard condition (B3), and from there the next steps were highly predictable. Of the steps out of that state, 70\% added \texttt{IndexError} (B4), 95\% of the following ones completed the arrangement with the return statement (B6), and 96\% of the students who reached that arrangement submitted it and failed. After a failed submission, the most common step (54\%) removed only the guard; the correct guard followed in 75\% of cases, and that arrangement passed in 97\%. The other route started with the correct guard (31\%) and led directly to a correct submission (67\%, 94\%, and 97\% at each step). A smaller branch combined the incorrect guard with \texttt{ValueError} (B3 and B5) and also ended in a failed submission (92\%).

Two kinds of state sit off the predictable routes. Students cleared the solution area 65 times, but never through a predictable step, so the reset state and the states around it are drawn in grey, off the predictable path. The map also shows two frequent but unpredictable moves after a failure. Fourteen students left the problem after a failed submission, which is 8\% of the steps that follow one, and the guard that returns \texttt{None} (B2) replaced the removed guard 14 times, failing again in 90\% of cases.

The two problems differ in where the most common route leads. On P1 it reaches a correct submission directly, with students placing the guard, the exception and the return in that order, and 82 of the 182 students who placed a block followed it from the empty solution area to the correct submission without a single extra step. The most common route on P3 is the one described above, which passes through a failed submission, and 39 of the 179 students followed it exactly. In both problems it is the single most frequent trajectory, taken by more students than any other sequence.

The four error categories behind those failures follow in turn.

\subsubsection{Return Instead of Raise}
\label{sec:return-throw}

Substituting a \texttt{return} statement for raising an exception recurred in all three problems (Table~\ref{tab:combined}). In P1, 42\% of students with incorrect attempts placed \texttt{if b == 0: return 0}, silently returning a value for a division by zero; in P2, 43\% placed \texttt{return 0} instead of raising \texttt{ValueError}, treating a parse failure as a silent exit; and in P3, 21\% placed a guard that returns \texttt{None}. In each case, students treated the error condition as a reason to leave the function early rather than as an event to signal to the caller.

\subsubsection{Wrong Exception Type}
\label{sec:exception-type}

Choosing an imprecise or wrong exception type was the most frequent pattern in P1 and P2. In P1, 76\% of students with incorrect attempts raised \texttt{RuntimeError} or \texttt{Exception} instead of \texttt{ZeroDivisionError}; in P2, 75\% caught \texttt{Exception} instead of \texttt{ValueError} or raised \texttt{RuntimeError}. In P3, 27\% raised \texttt{ValueError} instead of \texttt{IndexError}. In P1 and P2 this pattern combines two distractor blocks, and neither of them dominates on its own, accounting for 45\% and 26\% of P1's failed submissions and for 37\% and 24\% of P2's (Table~\ref{tab:failed}).

\subsubsection{Wrong Guard Condition}
\label{sec:guard}

The error that dominated P3 was neither of the above. Of the students with incorrect attempts, 89\% (110 of 123) chose the guard \texttt{if index < 0} over the correct bound \texttt{if index < -len(items)}, which is needed to reject out-of-range negative indices. It appears in 70\% of P3's failed submissions, more than three times the next block (Table~\ref{tab:failed}). P3 was also the hardest of the three problems, with 28.3\% of students submitting a correct first arrangement compared with 76.0\% in P1 and 56.6\% in P2 (Table~\ref{tab:combined}). Only P3 offered such a distractor, so whether the pattern generalizes to other boundary conditions is untested.

\subsubsection{Wrong Block Order}
\label{sec:control-flow}

Submitting correct blocks in the wrong position was most common in P2, affecting 58\% of students with incorrect attempts, against 24\% in P1 and 20\% in P3. P2 differs from the other two problems in more than one respect, since it is the only one with four slots and the only one whose solution nests a call inside a \texttt{try} and re-raises from the handler. With three problems we cannot say which of the two accounts for the difference.

\FloatBarrier
\subsection{RQ3: Common Difficulty Patterns Across Problems}
\label{sec:rq3}

Three of the four errors of Table~\ref{tab:combined} appear in all three problems and can be followed across them: choosing the wrong exception type, returning a value instead of raising an exception, and placing correct blocks in the wrong position. The fourth, the wrong condition, is left out because only P3 offers a guard distractor. We follow the 171 students who submitted on all three problems, identified by the pseudonym the platform assigns them, which is the same across the three, and count a student as making an error on a problem if at least one of their submissions there contained a distractor of that kind.

The question is whether making an error on the first problem says anything about the problems that follow. We therefore divided the students according to what they did on P1 and compared how often each group made the same error on P2 or P3 (Table~\ref{tab:recurrence}). The comparison group consists of students who did not make that error on P1 but had the same two later opportunities to make it.

Of the 27 students who chose the wrong exception type on P1, 63\% chose it again later, against 35\% of the 144 who had not (risk ratio 1.78, $p=0.010$ by Fisher's exact test, $p=0.020$ after Holm correction across the three errors). Of the 16 who returned a value instead of raising, 62\% did it again, against 25\% (2.48, $p=0.003$, corrected $p=0.010$). Of the 9 who misplaced correct blocks, 67\% did it again, against 27\% (2.45, $p=0.019$, corrected $p=0.020$).

All three differences remain statistically significant after correction, and none of the three confidence intervals includes equal risk. They run from 1.24 to 2.56 for the exception type, from 1.56 to 3.96 for returning a value, and from 1.45 to 4.15 for misplaced blocks. Taken together, a student who makes one of these errors on the first problem is between 1.2 and 4.2 times as likely to make it again on a later one, depending on the error and on where within each interval the true value falls. The intervals are wide because few students make each error on P1, namely 27, 16 and 9 respectively. The direction and the significance therefore hold, but the point estimates are imprecise, and the 67\% reported for misplaced blocks rests on 6 students.

The wrong exception type reappears mostly on P2, where 56\% of the students who made it on P1 made it again against 24\% of the rest, and evens out on P3, 22\% against 19\%. Returning a value instead of raising does the opposite, 19\% against 15\% on P2 and 50\% against 12\% on P3. Only misplaced blocks behave alike on both. We do not test the per-problem splits and some of their cells hold three students, but they show that which problem comes next matters as much as which error a student made before.

\begin{table}[!ht]
\caption{Whether an error made on P1 reappears on P2 or P3, among the 171 students who submitted on all three problems. \textit{Made it on P1} and \textit{Did not} split the students by what they did on P1, and the columns under each give how many of that group made the error on P2, on P3, and on either, which is a repetition for the first group and a first occurrence for the second. RR, $p$ and $p_\text{H}$ compare the two \textit{Either} columns, so the per-problem columns are descriptive and are not tested. RR = risk ratio between the two percentages; $p$ = Fisher's exact test; $p_\text{H}$ = the same $p$ after Holm's correction across the three errors.}
\label{tab:recurrence}
\footnotesize
\setlength{\tabcolsep}{5pt}
\begin{tabular}{@{}l rrrr rrrr rrr@{}}
\toprule
& \multicolumn{4}{c}{\textit{Made it on P1}} & \multicolumn{4}{c}{\textit{Did not}} \\
\cmidrule(lr){2-5}\cmidrule(lr){6-9}
\textbf{Error} & \textit{n} & \textbf{P2} & \textbf{P3} & \textbf{Either}
 & \textit{n} & \textbf{P2} & \textbf{P3} & \textbf{Either}
 & \textbf{RR} & $p$ & $p_\text{H}$ \\
\midrule
Wrong exception type      & 27 & 15 (56\%) & 6 (22\%) & 17 (63\%) & 144 & 34 (24\%) & 27 (19\%) & 51 (35\%) & 1.78 & 0.010 & 0.020 \\
Return instead of raise   & 16 &  3 (19\%) & 8 (50\%) & 10 (62\%) & 155 & 23 (15\%) & 18 (12\%) & 39 (25\%) & 2.48 & 0.003 & 0.010 \\
Correct blocks misplaced  &  9 &  3 (33\%) & 3 (33\%) &  6 (67\%) & 162 & 33 (20\%) & 22 (14\%) & 44 (27\%) & 2.45 & 0.019 & 0.020 \\
\bottomrule
\end{tabular}
\end{table}

\FloatBarrier
\subsection{RQ4: Students Who Never Fail}
\label{sec:rq4}

Twenty-seven students, 16\% of the 171 who submitted on all three problems, solved every problem at the first attempt. They are a different 27 from those of Table~\ref{tab:recurrence}, where the same number happened to make the wrong exception type on P1. Over the whole assignment their median is 113 seconds, 11 block placements and three submissions, where the three solutions together require 10 blocks and three submissions. Their sessions stay direct as the problems get harder. In their traces, the number who go from the empty solution area to a correct submission without a single removal rises from 15 of the 27 on P1 to 18 on P2 and 19 on P3. The time they spend does not follow the difficulty of the problem either, since their median on P3 is 16 seconds against 37 on P1 and 42 on P2.

\begin{figure}[!ht]
  \centering
  \includegraphics[width=\linewidth,alt={Three cohort predictability maps, one per problem, for the 27 students who solved every problem at the first attempt}]{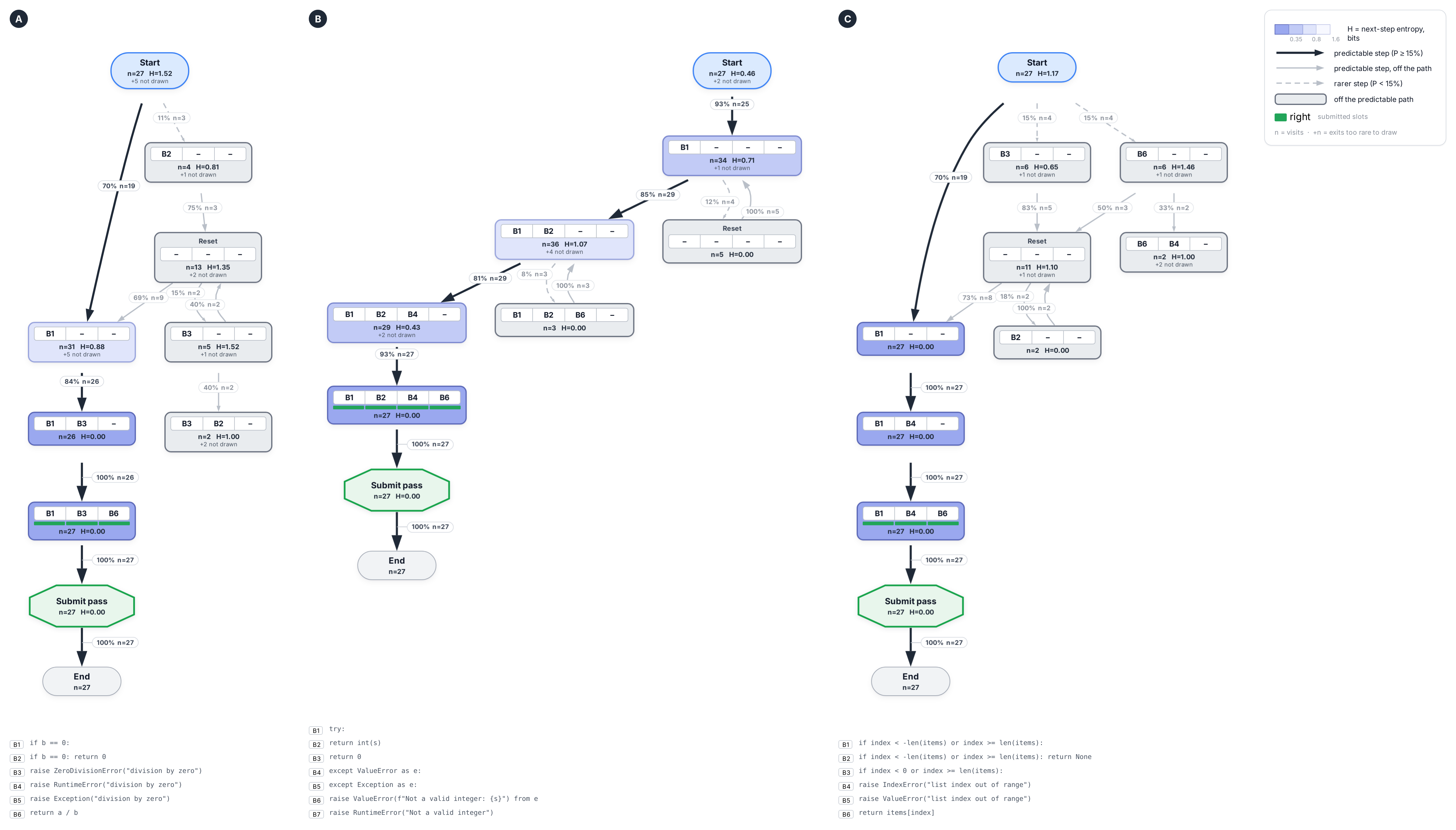}
  \caption{The 27 students who solved every problem at the first attempt, with the encodings of Figure~\ref{fig:cohort-map}. \textbf{A} is P1, \textbf{B} is P2 and \textbf{C} is P3. The thresholds are adapted to the smaller group, so a state needs three visits and a transition two occurrences to be drawn.}
  \Description{Three state maps side by side. In each of them a dark spine runs from a start node through two or three arrangement states to a passed submission, with the percentages of the steps along it close to 100. A few grey states sit to the side of the spine, among them the state in which the solution area is empty, and the arrows that leave them lead back to the spine or to the empty solution area rather than to a submission. A shared legend appears on the right and the block bank of each problem under its panel.}
  \label{fig:neverfail}
\end{figure}

Figure~\ref{fig:neverfail} draws the same kind of map for those 27 students, one panel per problem. On each problem their route from the empty solution area to a correct submission is a single dark spine, and the branching happens at the first block and almost nowhere else. In P1, 70\% of them opened with the correct guard, 84\% of the steps out of that state added the correct exception, and every student who reached the two following states took the same step. In P2 the opening is nearly fixed, since 93\% began with \texttt{try:} and the entropy of their first step is 0.46 bits. In P3, 70\% opened with the correct guard against 31\% of the cohort in Figure~\ref{fig:cohort-map}, and four of them opened with the guard that traps half the class.

The maps also show where this group's mistakes go. They do reach states off the route, with 13 visits to an empty solution area in P1 and 11 in P3, and they leave those states by erasing rather than by submitting. In P3, five of the six steps out of the incorrect guard cleared the solution area, and the sixth is too rare to be drawn. The 27 therefore do not avoid the distractors altogether, they remove them before submitting.

\FloatBarrier

Of the 171 students, 131 solved P1 at the first attempt (Table~\ref{tab:time} reports 133, counting two students who did not submit on all three), 83 of those were still without a failure after P2, and 27 after P3. Two thirds of them carried over to P2, and a third of those to P3.

Solving P1 at the first attempt carries over to P2, since 63\% of the 131 who did also solved P2 at the first attempt, against 38\% of the 40 who did not (risk ratio 1.69, $p=0.006$ by Fisher's exact test). It does not carry over to P3, where the two groups are almost indistinguishable, 29\% against 25\% (1.16, $p=0.69$). The effect is larger on P2 than on P3, but the difference between the two is not itself significant ($p=0.06$).

The difference between the two steps follows the difficulty of the problems (Table~\ref{tab:combined}). P2 is solved at the first attempt by 56.6\% of the students and P3 by 28.3\%, and P3 is the problem in which one distractor, the incorrect guard condition, carries 89\% of the students with a wrong submission. That distractor caught the students who had solved P1 immediately as often as the rest, since 93 of the 131, 71\%, did not solve P3 at the first attempt.

Solving the first problem without error is therefore informative about the second problem, and we cannot show that it is informative about the third. Membership of the group cannot be predicted from an earlier problem, but the traces identify it once the assignment is done, which is when an instructor would decide what to set these students next.

\section{Discussion}
\label{sec:discussion}

The overall aim of this work was to make the problem-solving processes of the students more visible to the instructors. The tool ran first in a Java pilot and then in a Python course with 176 students, whose submissions revealed four recurring errors (Table~\ref{tab:combined}) and whose event log made the trajectories behind them visible.

A student who chose the wrong exception type on P1 was 1.8 times as likely to choose it again later, and the two other errors we could follow gave 2.5 and 2.4 (Table~\ref{tab:recurrence}), so an instructor who sees one of these errors once can expect to see it again. Which later problem it reappears in depends on the problem rather than on the student, since the exception type came back mostly in P2 and returning a value instead of raising in P3, so the moment to address an error is set by what the assignment asks next. The intervention therefore belongs before the problem that will ask for the same concept again, rather than after the assignment is complete.

On every problem the students who never solved it spent about as long as those who solved it at the first attempt, yet the two groups were doing opposite things. Those who never solved it moved every 2.3 to 2.8 seconds and stopped, while those who solved it at the first attempt took 3.8 to 6.7 seconds over each move and submitted a correct arrangement (Table~\ref{tab:time}). A dashboard reporting only correctness and time on task would show them as two short sessions with different outcomes and offer no reason for the difference. The two groups also call for opposite responses, since those who stopped need to be reached before they give up on the topic, while those who solved at the first attempt are the ones an instructor would extend, once the traces identify them at the end of the assignment (Section~\ref{sec:rq4}).

The same traces show how cheaply a failed submission can be repaired. On P3, the most common move after a failure removed only the guard, the correct guard followed in 75\% of those cases, and that arrangement passed in 97\% (Figure~\ref{fig:cohort-map}). With three candidate guards in the bank and a red marker on the offending block, a student can reach the correct arrangement by elimination without ever reasoning about why the bound needs \texttt{-len(items)}. Per-block feedback therefore buys a fast recovery at the price of turning a conceptual question into a search, and the narrower the set of alternatives, the more that search pays off. A correct final arrangement is not by itself evidence that the concept behind it was understood, and only the trace distinguishes the two cases. Whether the students who repaired this way understood the bound afterwards is not something the traces can tell us, and isomorphic variants of the same problem~\cite{tilantera2025traces} could be one way to find out.

Knowing which distractor carries the failures also feeds back into problem design. A distractor that almost nobody selects adds little, while one that carries most of the failures, such as the incorrect guard in P3, marks a concept close enough to the correct answer to be confusable. Trace data makes that distinction measurable rather than a matter of intuition, and it suggests a way to calibrate difficulty. The next version of a problem can be built around the distractors that proved confusable, retiring those that did not. It also raises a question this study cannot answer, namely what makes a distractor confusable in the first place, whether it is proximity to the correct concept, surface similarity of the code, or familiarity from earlier material.

\subsection{Instructor Perspectives}
\label{sec:perspectives}

The class-level aggregation (panel~J, Figure~\ref{fig:overview}) is the feature the pilot instructor singled out, having used the generated slides to go over common mistakes in class.

The main deployment had two instructors, one of whom used the dashboard and described the experience as follows:

\begin{tcolorbox}[
  breakable,
  colback=gray!8, colframe=gray!30,
  arc=2pt, boxrule=0.4pt,
  left=5pt, right=5pt, top=2pt, bottom=2pt
]
\footnotesize
From an instructor perspective, the dashboard was useful because it made students' intermediate solution attempts visible rather than showing only the final outcome. The color-coded attempt history made it easy to see which lines were correct, which were merely out of place, and which were incorrect distractors, helping me identify whether a student was close to the solution or repeatedly applying the same misconception. I found the AI-assisted analysis useful as a starting point for interpreting these patterns, but I would still want to review the traces of students before deciding what feedback or intervention is appropriate.
\end{tcolorbox}

The same instructor used both the per-student analysis and the class-level slides, and described the per-student view as follows:

\begin{tcolorbox}[
  breakable,
  colback=gray!8, colframe=gray!30,
  arc=2pt, boxrule=0.4pt,
  left=5pt, right=5pt, top=2pt, bottom=2pt
]
\footnotesize
I think this provides a nice way to see how an individual student worked through a problem. Being able to follow their problem-solving process, rather than just see the final submission, is particularly useful. In a class this large, I probably wouldn't have time to go through every student individually. However, I could see this being useful when a student asks for help with a particular problem. I could open their submission and walk through the process together with them to discuss the difficulties they encountered.
\end{tcolorbox}

The view is therefore used at the point where a student already asks for help, rather than as a review task added to a course of this size. The instructor treated the trace, not the analysis of it, as what they would act on.

The class-level deck would be used in a lecture following the activity to review common misconceptions, with minor modifications. Because the deck is presented to a class, its contents must be verifiable before use. The current interface reports the extracted difficulties without the submissions behind them. Showing that evidence alongside each item, and restricting the generated examples to the code of the assignment, would let an instructor verify a deck before presenting it.


\section{Limitations}
\label{sec:limitations}

Several limitations apply. First, all reported results come from a single course, institution, and topic, so we cannot say whether the patterns generalize to other programming courses or topics. Two of the four error categories were already the dominant failures in the upper-division Java pilot, but it differed from the main deployment in too many respects to serve as a comparison. Second, access was anonymous, so we have no information on students' backgrounds and cannot link their traces to other course activities. Third, the course was self-paced, and the students who attempted the assignment may not represent the full enrollment. Fourth, both deployments were single assignments in single courses, so we have no evidence on how the tool behaves at a larger scale or in courses of other kinds.

The method also bounds what can be observed. Traces record which block a student placed and when, not why. The same distractor choice can reflect a misconception, a guess, or a slip, and without verbal data we cannot tell them apart. Students could only reveal the errors that the distractors offered, and the three problems covered a single topic, so other misconceptions about exception handling remain unobserved. The per-block feedback also shapes what students do after a failed submission, because it marks exactly which blocks to change; the trajectories we report may therefore not transfer to tools with binary or delayed feedback. Because students worked unsupervised and at their own pace, we cannot know whether they consulted others or external resources, and long pauses between events may reflect work outside the tool.

Finally, the trajectory analysis is largely descriptive. The Markov model also assumes that the next state depends only on the current one, an assumption we adopt from \citet{kumar2023markov} without testing it. A student who reaches an arrangement after several erasures may well continue differently from one who reaches the same arrangement directly. Solution states were reconstructed from interaction events and checked against the recorded submissions; in 4 of 540 trajectories the reconstruction was inconsistent with them, most likely because of work carried over from an earlier session. The thresholds of the cohort map are display decisions. Only two instructors used the dashboard, so their accounts are not evidence that it is useful to instructors in general. We show that the traces make difficulties visible to instructors, not that seeing them changed what instructors did or what students learned; demonstrating that would require comparing courses in which instructors did and did not have the dashboard. Distinguishing misconceptions from guesses would require verbal data that these traces do not contain.

\section{Future Work}
\label{sec:future}

Six extensions follow directly from the findings. First, replicating the deployment in other programming courses and topics would test whether these patterns generalize. Second, isomorphic variants~\cite{tilantera2025traces} would separate understanding from accidental correctness. Third, automated puzzles from recurring distractor selections~\cite{hou2024codetailor} would target these patterns directly. Fourth, real-time alerts on repeated failed arrangements~\cite{price2017isnap} would enable earlier intervention. Fifth, an in-dashboard state-transition graph would surface cycles and regressions automatically. Lastly, analyzing the quality of the AI-generated reports about common errors systematically would be important for quality assurance.

\begin{acks}
This material is based upon work supported by the U.S. National Science Foundation under Grant No. 2225373. Any opinions, findings,
and conclusions or recommendations expressed in this material are
those of the author(s) and do not necessarily reflect the views of
the National Science Foundation.
\end{acks}


\bibliographystyle{ACM-Reference-Format}
\bibliography{references}


\end{document}